\documentclass[a4paper,11pt]{article}
\usepackage{latexsym}
\usepackage{amsmath}
\usepackage{amssymb}
\usepackage{graphicx}
\usepackage{epsfig,amssymb,amsmath,graphicx,caption,subcaption,
verbatim,hyperref,xcolor,ulem,epstopdf,psfrag,pstool,braket,array,
enumerate,wrapfig,tabularx,setspace,geometry}
\usepackage{placeins}
\newcommand{\code}[1]{\texttt{#1}}

\newcounter{mnotecount}[section]
\renewcommand{\themnotecount}{\thesection.\arabic{mnotecount}}

\newcommand{\mnote}[1]
{\protect{\stepcounter{mnotecount}}$^{\mbox{\footnotesize
$
\bullet$\themnotecount}}$ \marginpar{
\raggedright\tiny\it 
$\!\!\!\!\!\!\,\bullet$\themnotecount: #1} }

\def\be{\begin{equation}}
\def\ee{\end{equation}}

\def\theequation{\thesection.\arabic{equation}}

\def\p{\partial}

\def\theequation{\thesection.\arabic{equation}}

\begin{document}
\begin{center}
{{\bf  \Large  Near--extremal gravitational collapse in 4+1 dimensions: Schwarzschild--de--Sitter space}} 
\\[4mm]
{{\bf \large Maciej Dunajski}\footnote{m.dunajski@damtp.cam.ac.uk}\\
Department of Applied Mathematics and Theoretical Physics\\ 
University of Cambridge\\ Wilberforce Road, Cambridge CB3 0WA, UK.\\
\vskip2pt
and\\
Faculty of Physics,  University of Warsaw\\
 Pasteura 5, 02-093 Warsaw, Poland.\\
\vskip2pt
{\bf\large Sebastian J. Szybka}\footnote{sebastian.szybka@uj.edu.pl}\\ Astronomical Observatory,
Jagiellonian University\\
Krak\'ow, Poland.}
\vskip5mm
{7 July, 2026}
	\abstract{We numerically study a formation of near extremal horizons from a gravitational collapse of radially symmetric gravitational waves in $4+1$ dimensions within the framework of pure Einstein gravity with positive cosmological constant. Evolution of a regular initial data with cosmological horizon leads to a formation of a black hole with  mass exceeding $99\%$ of the extremal value corresponding to the black hole and cosmological horizons coinciding.

We demonstrate how our results fit within the framework of characteristic gluing, and present some evidence that the third law of black hole thermodynamics may not hold in the cosmological context, where the extremality corresponds to the maximal mass of the Schwarzschild black hole in de--Sitter space.}
\end{center}

\section{Introduction}
Extremal black holes are characterised by vanishing surface gravity of their Killing horizon. They belong
to boundaries of stationary black holes parameter spaces: maximally spinning in the Kerr family, maximally charged in the Reissner-- Nordstr\"om (RN) family, or  maximally large in the Schwarzschild--de Sitter family.
The techniques used to study the subextremal black holes do not extend to extremal cases, 
and new techniques have been developed to study dynamical problems \cite{Dafermosessay}
as well as the rigidity theorems of extremal horizons \cite{DL, Col}.

While the astrophysical Kerr black holes can be close to extremal,  the angular momentum to mass ratio
can not exceed $0.998$ if a black hole gains its spin in an accretion process \cite{thorne}.
The extremal black holes, where this ratio is $1$,  are believed not to exist in Nature. On the theoretical side their existence has been ruled out by the {\it  third law of black hole thermodynamics} \cite{Hawking, Israel} which states that  a subextremal black hole can not become extremal in finite time evolving from regular initial data.
In a recent paper of Kehle and Unger 
\cite{KU} this law has been disproven: there exists asymptotically flat spherically symmetric data for the 
Einstein--Maxwell charged scalar field equation
which
results in a formation of an extremal RN horizon. A reason for incorporating the scalar field is that the asymptotically flat and spherically symmetric pure Einstein--Maxwell space--time in 3+1 dimensions must be static, so there is no gravitational collapse of spherically symmetric initial data. 

In the current  paper we explore a possible third law violation in the 4+1 dimensional pure Einstein gravity with positive cosmological constant, where the Birkhoff theorem can be evaded by a modification of the Bizo\'n--Chmaj--Schmidt (BCS) ansatz \cite{BCS_paper}, and the extremality parameter is taken to be a positive cosmological constant.  In their  remarkable paper \cite{BCS_paper} BCS introduce a squashing factor on the three--sphere, measured by the function $B=B(r, t)$. This preserves the radial symmetry, but allows for the time dependence. In the case analysed in \cite{BCS_paper} the dynamics of $B$ admits a family of attractors with $B=0$ parametrised by a non---negative number $m$ proportional  to the total ADM mass. Depending on the size of initial data the end point of the evolution is the Minkowski space
with $m=0$, or the 4+1 dimensional Schwarzschild solution with $m>0$.

 Our setup involves 
gravitational waves inside the cosmological horizon, with regular initial data with no black hole horizon.  During the evolution the waves can radiate away thus leading to a dispersion to de--Sitter, or to a formation of a Schwarzschild--de Sitter black hole which is what we focus on. In 
\S\ref{sectionBCS} and \S\ref{nullcorsS} we introduce a modification of the BCS ansatz in the presence of a cosmological constant. In \S\ref{section_numerics} we develop and implement a numerical scheme to analyse black hole formations within the BCS ansatz starting from a regular initial data which is asymptotically de--Sitter.  For a fixed value of the cosmological constant $\Lambda$, and depending on the size of the initial data, the resulting black hole can be small, with the mass  $M=0.0036 M_{\max}$ or near extremal with $M=0.99 M_{\max}$, where $M_{\max}=\frac{3}{2\Lambda}$ is the maximal mass corresponding to the extreme Schwarzschild--de-Sitter solution. In 
\S\ref{section_gluing} we put our results into the framework of the characteristic gluing originally used in \cite{KU} to disprove the 3rd law.
In the Appendix we present the linear asymptotic analysis of 
settling to a Schwarzschild--de--Sitter black hole dominated by quasi--normal modes,  and obtain an analytical formula for quasi--normal frequencies in the near--extremal case.

\section{The BCS ansatz}
\label{sectionBCS}
Let $(M, g)$ be a Lorentzian 4+1--dimensional manifold with a metric of the form \cite{BCS_paper}
\be
\label{BCSmetric}
g=-Ae^{-2\delta}dt^2+A^{-1}dr^2+\frac{1}{4}
r^2\Big(e^{2B}({\sigma_1}^2+{\sigma_2}^2)
+e^{-4B}{\sigma_3}^2\Big)
\ee
where $A, B, \delta$ are functions of $(r, t)$. The metric (\ref{BCSmetric}) admits an isometric action of 
$SU(2)\times U(1)$, where the $U(1)$ action commutes with that of $SU(2)$. The group $SU(2)$ acts
on $M$ with $3$--dimensional orbits, and  the Einstein  equations with cosmological constant  $\Lambda$ 
reduce to four PDEs for  three functions $(A, B, \delta)$ a $1+1$ dimensional quotient $\Sigma$ of $M$
\begin{subequations}
\begin{align}
\label{em1}
A'&=-\frac{2A}{r}+\frac{1}{3r}(8e^{-2B}-2e^{-8B})-2r(e^{2\delta}A^{-1}\dot{B}^2
+A{B'}^2) -\frac{2}{3}\Lambda r\\
\label{em2}
\delta'&=-2r(e^{2\delta}A^{-2}{\dot{B}}^2+{B'}^2) \\
\label{em3}
0&=\Big(e^{\delta}A^{-1}r^3\dot{B}\Big) ^{\cdot}-
(e^{-\delta} Ar^3B')'+
\frac{4}{3}e^{-\delta}r(e^{-2B}-e^{-8B})\\
\label{em4}
\dot{A}&=-4rA\dot{B}B' 
\end{align}
\end{subequations}
where $A'=\p_r A, \dot{A}=\p_t A$ etc. 
There exists a two--parameter family of static solutions to (\ref{em1})--(\ref{em4}) given by
$B=0, \delta=0$ and
\begin{eqnarray}
\label{static}
A=A_0&=&1-\frac{m}{r^2}
- \frac{1}{6}\Lambda r^2.\nonumber\\
&=&\frac{\Lambda}{6r^2}({r_c}^2-r^2)(r^2-{r_s}^2),
\end{eqnarray}
where the black hole
horizon $r_s$ and the cosmological horizon $r_c$ are given by
\be
\label{rsrc}
r_s=\sqrt{\frac{3}{\Lambda}\Big(1-\sqrt{1-\frac{2}{3}\Lambda m}\Big)}, \quad
r_c=\sqrt{\frac{3}{\Lambda}\Big(1 +\sqrt{1-\frac{2}{3}\Lambda m}\Big)}
\ee
and $\Lambda m\leq \frac{3}{2}$.
These solutions, which are natural generalisations  of the Kottler metric \cite{Kottler} to $4+1$ dimensions, admit a symmetry enhancement from $SU(2)\times U(1)$ to $SO(4)$.
The corresponding metric describes the region of $4+1$ dimensional Schwarzschild--de--Sitter (SdS) black hole in the static patch\footnote{The following coordinate system generalises (with dimension--dependent differences) the coordinates
of 3+1 dimensional SdS in \cite{GP}. 
\be
\label{GPmetric}
g=-\Big(\frac{1-\frac{m}{4R^2} e^{-2T/a}}{1+\frac{m}{4R^2} e^{-2T/a}}\Big)^2dT^2
+\Big( 1+\frac{m}{4R^2}e^{-2T/a}\Big)^2 e^{2T/a}
\Big(dR^2+\frac{1}{4}R^2({\sigma_1}^2+
{\sigma_2}^2+{\sigma_3}^2)\Big)
\ee
where 
\[
R=\frac{1}{2}(r+\sqrt{r^2-m})e^{-T/a}, \quad T=t-\int\Big(
\frac{r^{2} \left(-2 \sqrt{r^{2}-m}\, r^{2}-2 r^{3}+\sqrt{r^{2}-m}\, m +2 m r \right)}{a \left(-r^{2}+m \right) \left(2 r \sqrt{r^{2}-m}+2 r^{2}-m \right)}\frac{1}{A_0}\Big)dr, \quad \Lambda=\frac{6}{a^2}.
\]
If $m=0$ then (\ref{GPmetric}) reduces to $4+1$ de Sitter in the flat slicing coordinates which cover a half of the conformal diagram, so twice the size of the static patch. If $\Lambda=0$ then (\ref{GPmetric}) reduces to the isotropic coordinates for the $(4+1)$--dimensional Schwarzschild metric. Despite its radial symmetry 
the solution (\ref{GPmetric}) does not show in the BCS ansatz as an exact time dependent solution, as putting it in the form (\ref{BCSmetric})  requires a coordinate transformation to SdS coordinates in the static patch.}.
A more general, tri--axial, version of (\ref{BCSmetric}) which only admits $SU(2)$ as isometry group
has been analysed in \cite{DR, Szybka} when $\Lambda=0$. We shall only consider the biaxial ansatz in this paper.
\section{Double null coordinates and a horizon formation}
\label{nullcorsS}
While the static patch $(r, t)$ underlying  (\ref{BCSmetric}) is convenient in the Cauchy problem, and its  numerical implementation in \S\ref{section_numerics}, the discussion of the horizon formation (both event, and apparent) as well as the characteristic initial value problem in \S\ref{section_gluing} requires putting the 
BCS ansatz in the double--null coordinate form \cite{DR}
\be
\label{gEF}
g=-e^{2F}dudv +\frac{1}{4}
r^2\Big(e^{2B}({\sigma_1}^2+{\sigma_2}^2)
+e^{-4B}{\sigma_3}^2\Big)
\ee
where now $F=F(u, v),\, r=r(u, v),\, B=B(u, v)$. 
The computation
of the Einstein tensor of (\ref{gEF})
 leads to five 2nd order PDEs,
but  only the following four
\begin{subequations}
\begin{align}
\label{emmm1}
r_{vv} &=2r_v F_v-2r{B_v}^2\\
\label{emmm2}
r_{uu} &=2r_uF_u-2r{B_u}^2\\
\label{emmm3}
r_{uv} &=\frac{\Lambda re^{2F}}{6}-\frac{2}{r}r_ur_v
+\frac{e^{2F}}{6r}(e^{-8B}-4e^{-2B})
  \\
\label{emmm4}
B_{uv} &= \frac{e^{2F}}{3r^2}( e^{-8B}-e^{-2B} )  
-\frac{3}{2r}(r_vB_u+r_uB_v)
\end{align}
\end{subequations}
are independent. Indeed,  subtracting  $B_v$ times the $v$ derivative
of (\ref{emmm1}) from $B_u$ times the $u$ derivative
of (\ref{emmm2}), eliminating the third derivatives using the derivatives of (\ref{emmm3}) and finally
using (\ref{emmm1}),(\ref{emmm2}), (\ref{emmm3}) to eliminate
$(r_{uu}, r_{uv}, r_{vv})$ from the resulting equation
yields the final PDE resulting from the Einstein equations
\be
\label{lasteeq}
F_{uv}=\frac{e^{2F}}{r^2}\Big(e^{-2B}-\frac{1}{4}
e^{-8B}\Big)+\frac{3}{r^2}r_ur_v-3B_uB_v-\frac{\Lambda}{12}e^{2F}.
\ee
In 4+1 dimensions, and with radial symmetry, the Hawking mass reduces to the 
Misner-Sharp mass
\be
\label{MSmass}
M=r^{2}(1-{|\nabla r|_g}^2)
\ee
with the total mass $M_{\infty}=\frac{3\pi}{8} \lim_{r\rightarrow \infty} M$ for asymptotically flat
space--times.
In the presence of the cosmological constant $\Lambda$ we shall use the renormalised MS 
mass
\begin{eqnarray}
\label{MSmassL}
M_{\Lambda}&=& r^{2}(1-{|\nabla r|_g}^2-\frac{1}{6}\Lambda r^2)\nonumber\\
&=&r^2(1+4e^{-2F}r_u r_v-\frac{1}{6}\Lambda r^2)
\end{eqnarray}
which ensures that $M_{\Lambda}=0$ for the pure de--Sitter space\footnote{If we instead used
(\ref{MSmass})
with  ${|\nabla r|_g}^2=1-m/r^2-\Lambda r^2/6$
then the mass of the spherical shell 
of radius $r$ in the SdS space would be [with the $3\pi/8$ factor re--introduced]
\be
\label{nonmodifiedsds}
M=\frac{3\pi}{8}\Big(m+\frac{\Lambda}{6} r^4\Big).
\ee
The quartic term could account for the non--relativistic Hooke law \cite{Dun}, but with the opposite sign in the force term.}.  
The modified Misner--Sharp mass (\ref{MSmassL}) 
satisfies monotonicity properties resulting from 
\begin{eqnarray*}
\p_u M_{\Lambda} &=&\frac{2}{3}rr_u(3+e^{-8B}-4e^{-2B})-8e^{-2F}{B_u}^2r^3r_v\\
\p_v M_{\Lambda} &=&\frac{2}{3}rr_v(3+e^{-8B}-4e^{-2B})-8e^{-2F}{B_v}^2r^3r_u.
\end{eqnarray*}
The function $3+e^{-8B}-4e^{-2B}$ is non--negative with a single minimum at $B=0$. Therefore the monotonicity properties of the Misner-Sharp mass depend on the signs of the expansion scalars
$\theta_-= 3e^{-2F}r^{-1}r_u$ and $\theta_+=3e^{-2F}r^{-1}r_v$. A sphere of radius $r$ is trapped if $\theta_-,\, \theta_+$ are negative and so
\be
\label{MScontrol}
\frac{M_{\Lambda}}{r^2}+\frac{1}{6}\Lambda r^2-1>0,
\ee
therefore the formation of black holes is controlled by the LHS of this inequality.
\subsection{Horizons}
The event horizon separates outgoing null geodesics which reach ${\mathcal I}_+$
from those which do not. This definition is {\it teleological} as it requires the knowledge of the entire future of the space--time.
To account for an observer--dependent cosmological horizon consider a time--like curve  $\gamma$
of infinite proper length  (this is the worldine of an observer which does not fall into a black hole)
and define the future event horizon of ${\mathcal H}^+(\gamma)= \partial(J^{-}(\gamma)) $ to be the boundary of causal past of the future end--point of
$\gamma$ on ${\mathcal I}^+$. This splits into the union of ${\mathcal H}_{BH}=\partial (J^{-}({\mathcal I}^+))
$ which is observer independent and its set complement, the cosmological horizon, which is 
the past achronal boundary of $\gamma$ excluding the black hole horizon. It exists already in the pure de--Sitter space as a consequence of the infinities
${\mathcal I}^{\pm}$ being space--like.

In the numerical computations we shall therefore focus on the formation
of the apparent horizon,
 which is the outermost marginally trapped surface where $\theta_+=0$ and $\theta_-<0$. While for sub-extremal black holes $\theta_+$ flips sign passing through this horizon, for extremal black holes it attains a minimum  at $0$ (see \cite{Israel} for a definition of a dynamical extremal black hole).
 The  location of the apparent horizon is defined by the vanishing of the
 function $A(r, t)$  in (\ref{BCSmetric}) (there is always one zero corresponding
to the cosmological horizon), or equivalently vanishing of $e^{2F}$ in (\ref{gEF}). In the cosmological collapse of (\ref{BCSmetric}) the event horizon will form first with some null geodesics seemingly able to escape to ${\mathcal I}_+$ only to be trapped later. In a collapse to the static Schwarzschild-de-Sitter black hole the coordinate $u$ at which the apparent horizon intersects ${\mathcal I}^+$ defines the event horizon.
Expressing the  apparent horizon condition $A(r, t)=0$ as $u=f(v)$ gives the location of the event horizon as $u_{EH}=\lim_{v\rightarrow \infty} f(v)$.
which holds if a dynamical black hole settles to a stationary one.
\section{Numerical results}
\label{section_numerics} 
In the numerical implementation it is convenient to set  $P=re^{\delta} A^{-1}\dot{B}$ in terms of which the equations (\ref{em1})--(\ref{em3}) form a first order system for $A$, $B$, $\delta$, $P$. The last equation \eqref{em4} is redundant: we use it  to monitor the  accuracy of the code, and not to advance $A$ in time.

We impose the following  boundary conditions
\begin{equation}
        A(t,0)=1\;, \quad B(t,0)=0\;, \quad \delta(t,0)=0\;
\end{equation}
at the centre which, in the numerical implementation,  is approximated by $r=10^{-6}$. These conditions 
give $g_{tt}=A e^{-2\delta}=1$ at the centre, fixing $t$ to coincide with a proper time of a stationary observer there.  We prescribe initial data at $t=0$ by
\begin{equation}\label{initial}
	B(0,r)=b_0 r^2 e^{-\left(\frac{r-r_0}{s}\right)^{4}}\,, \quad P(0,r)=-2\,v_0\,r\,B(0,r)\left(\frac{1}{r}-\frac{2}{s}\left(\frac{r-r_0}{s}\right)^3\right)\;,
\end{equation}
where $b_0$, $r_0$, $s$, $v_0$ are  parameters varied in the simulations. To motivate the choice of $P(0,r)$, consider  another pulse which propagates rigidly: $\hat B(t,r) = B(0, r - v_0 t)$. Increasing $t$ shifts the profile with coordinate velocity $v_0$. The pulse is outgoing if $v_0>0$ and ingoing if $v_0<0$. Differentiating yields
\begin{equation}
	\partial_t \hat B(t,r)|_{t=0}=-v_0\,\partial_r B(0,r)\;,
\end{equation}
and we choose this condition at $t=0$, namely, we set $\partial_t B(t,r)|_{t=0}=\partial_t \hat B(t,r)|_{t=0}$. We have $P=re^{\delta} A^{-1} {\dot B}$, but $\delta$ and $A$ are not known because we have not solved the constraint yet. Therefore, we approximate $e^{\delta(0,r)} A^{-1}(0,r)\approx 1$ at $t=0$ in the definition of $P(0,r)$. We set
\begin{equation}
	P(0,r)=-v_0\, r\,\partial_r B(0,r)\;,
\end{equation}
which coincides with  \eqref{initial}.
The coordinate speed of light for a radial signal is given by $v_c=dr/dt=A e^{-\delta}$ and the code verifies that $|v_0|<v_c(r)$ at $t=0$ over the pulse support ($|B(0,r)|>10^{-6}\,max|B(0,r)|$) after the constraint are solved. As $v_c$ vanishes at the cosmological horizon,  the outgoing perturbations are frozen there  which influences convergence as explained in Appendix B.

The equations (\ref{em1}--\ref{em4}) are solved as follows: From the initial data we solve the constraints for $\delta(0,r)$ and $A(0,r)$, then advance $B$ and $P$ from $t = 0$ to $t = \Delta t$ by the implicit Crank–Nicolson 
  discretisation of the evolution equations, solved by Newton iteration. After each time step the constraints are re-solved for the updated $\delta$ and $A$, and the      
  procedure repeats. The equation for $A'$ contains $A$ on the right hand side, so after discretization we solve an algebraic equation for $A$ at each time step. The equation involving $\delta'$ is also solved algebraically. The evolution equation for $A$ is only used to test consistency. 
 
The code is implemented in C++. We have verified its convergence (see Appendix B) . The spatial discretization
is second-order accurate. The temporal discretization is first-order accurate for fast-varying
$A$ because $A$ is lagged. If $A$ does not vary much, the effective temporal discretization
is second-order accurate. The grid is uniform in $r$ and has $n=44800$ nodes with 
$\Delta r = (r_{max} - r_{min}) /(n-1)$.
We have taken $r_{min}=10^{-6}$ for all sets of data
and different values for $r_{max}$ (see below) depending on the size of data.
In all cases the grid reaches beyond the cosmological horizon.

We have evolved three sets of initial data. The first set is a small perturbation of de--Sitter space, the second set is a larger perturbation which leads to a formation of a small black hole in a spirit of a critical collapse \cite{BCS_paper}. The third set is a large perturbation which leads to a formation of a near--extremal black hole.  We should stress that the size of the initial data (and therefore the resulting black hole) 
depends not only on the amplitude $b_0$, but also on $r_0$ which defines  the distance between the bump function and the cosmological horizon. A small amplitude perturbation near this horizon can have, on our grid,  a greater effect than a large amplitude perturbation near the centre.
The results are summarised in the next three subsections.

\begin{figure}[t]
    \centering                                                                                                                                                             
    \includegraphics[scale=0.6]{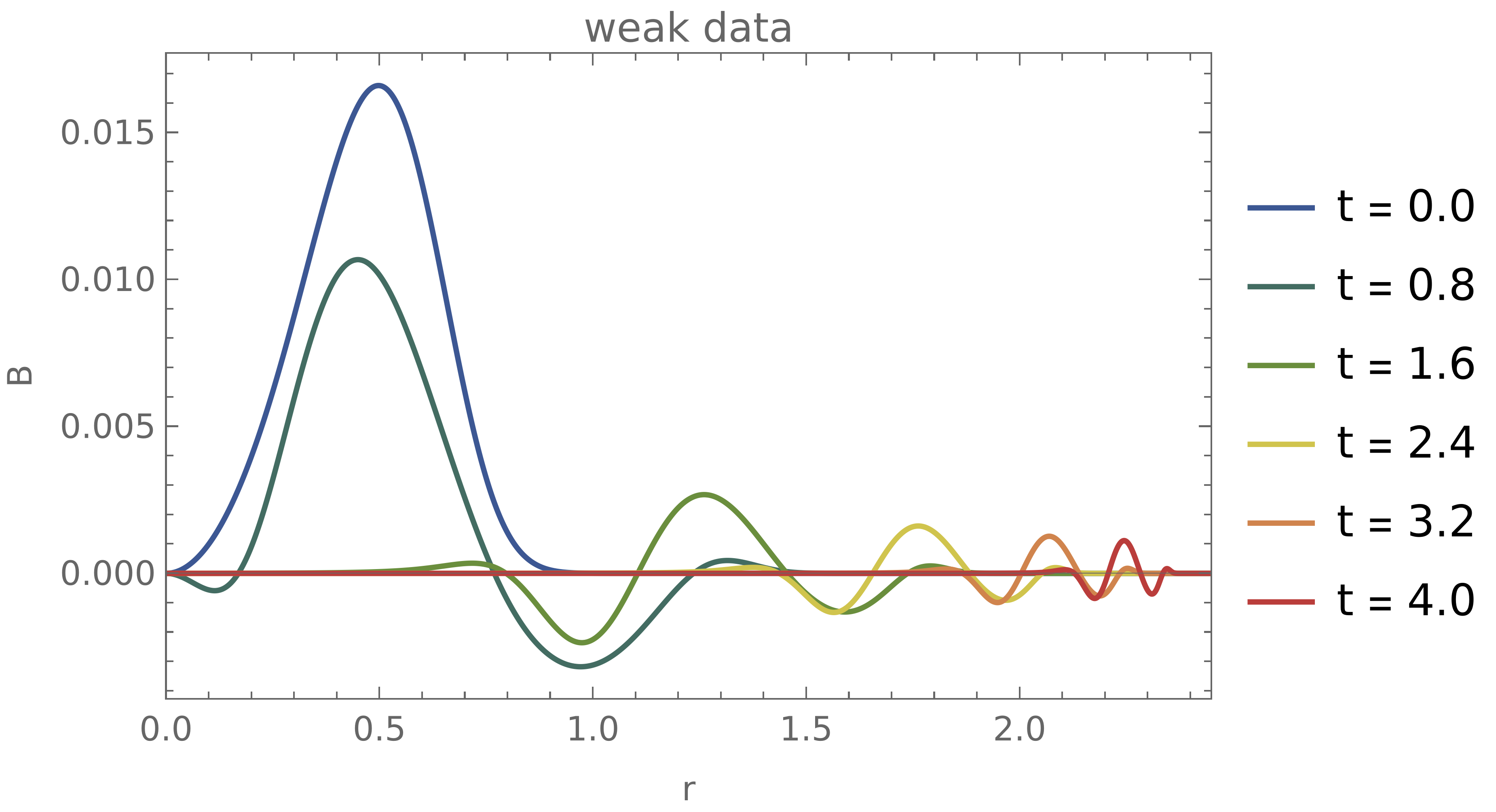}                     
    \caption{The squashing function $B$ for weak initial data.}                                                                                          
    \label{fig:Bw}                                                               
  \end{figure} 

\begin{figure}[h]
    \centering                                                                                                                                                             
    \includegraphics[scale=0.6]{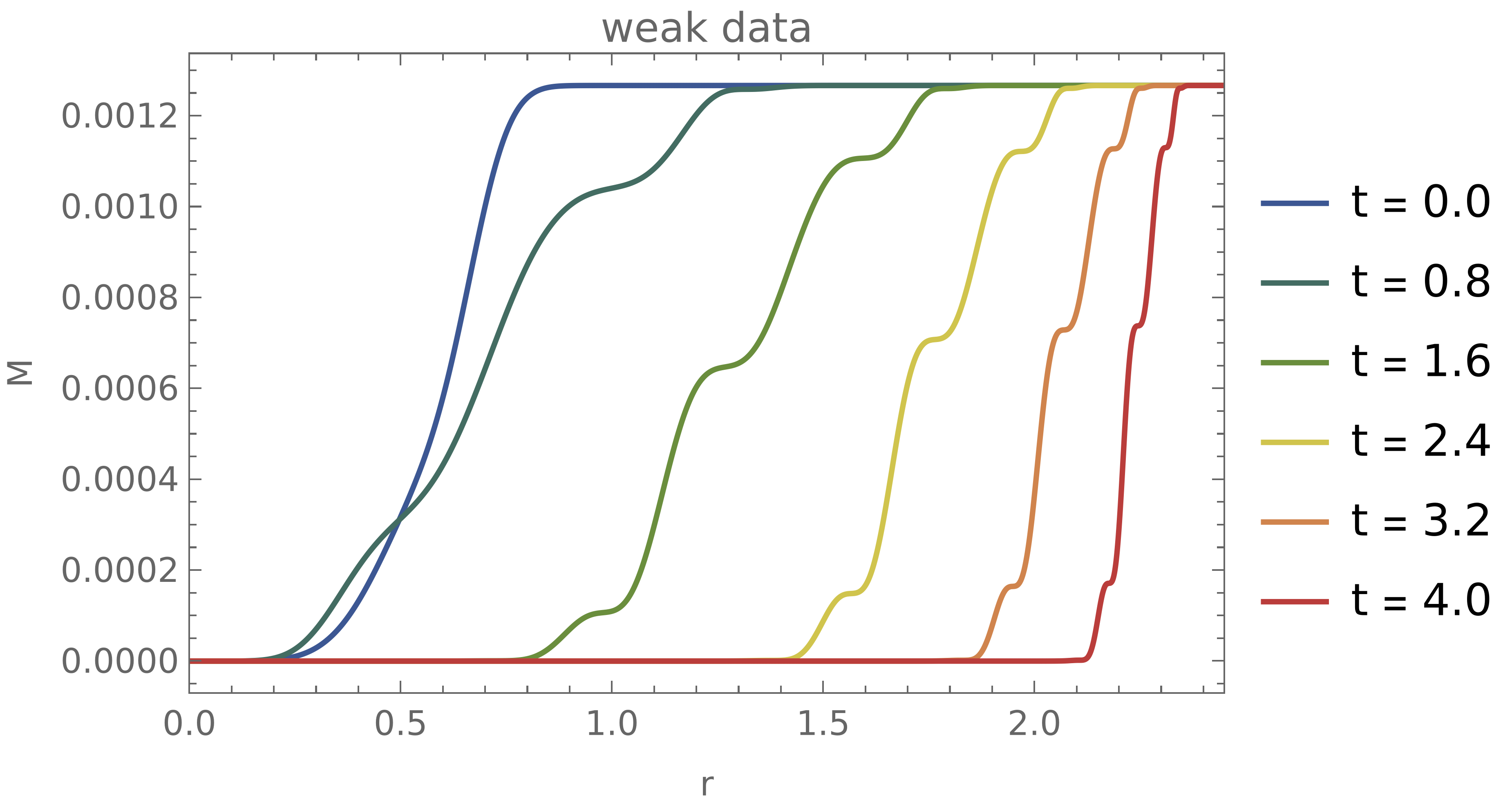}                     
    \caption{The Hawking mass $M$ for weak initial data.}                                                                                          
    \label{fig:Mw}                                                               
  \end{figure} 

\subsection{Weak data}

The evolution of weak initial data with $b_0=0.1$, $r_0=0.1$, $s=0.5$, $v_0=-0.5$, $\Delta t=2\cdot 10^{-6}$, $\Lambda=1$,
$r_{max}=2.5,
\Delta r=0.0000558048$
is shown in Fig.\ \ref{fig:Bw}. The initial data gives a small ingoing perturbation of de Sitter space at $t=0$, and the solution settles back to de Sitter after the perturbation (reflected from the center) propagate outwards. The coordinate system is singular at the cosmological horizon $r_c=2.449$ where the coordinate speed of light  of a radial signal vanishes, so we cannot follow radiation through the cosmological horizon. The time evolution of the Hawking mass $M$ shown in Fig.\ \ref{fig:Mw}, indicates that the energy is radiated away from the centre.

\subsection{Intermediate data}

\begin{figure}[p]

    \centering 
    \begin{minipage}{\textwidth}
    \centering                                                                                                                                                             
    \includegraphics[scale=0.6]{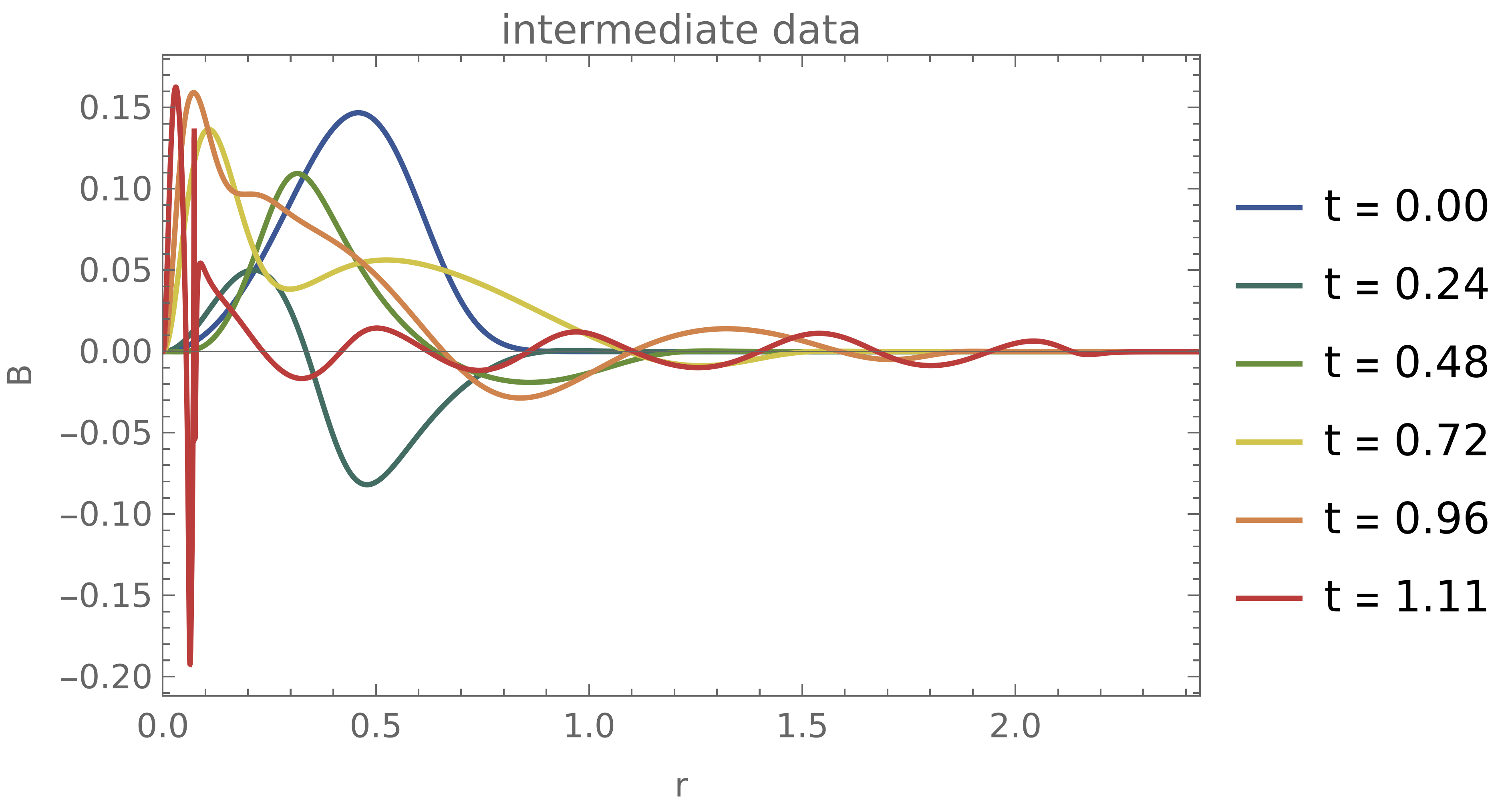}                     
    \caption{The squashing function $B$ for intermediate initial data.}                                                                                          
    \label{fig:Bi}                                                               
    \end{minipage}

	\vspace{1cm}

    \begin{minipage}{\textwidth}
    \centering                                                                                                                                                             
    \includegraphics[scale=0.6]{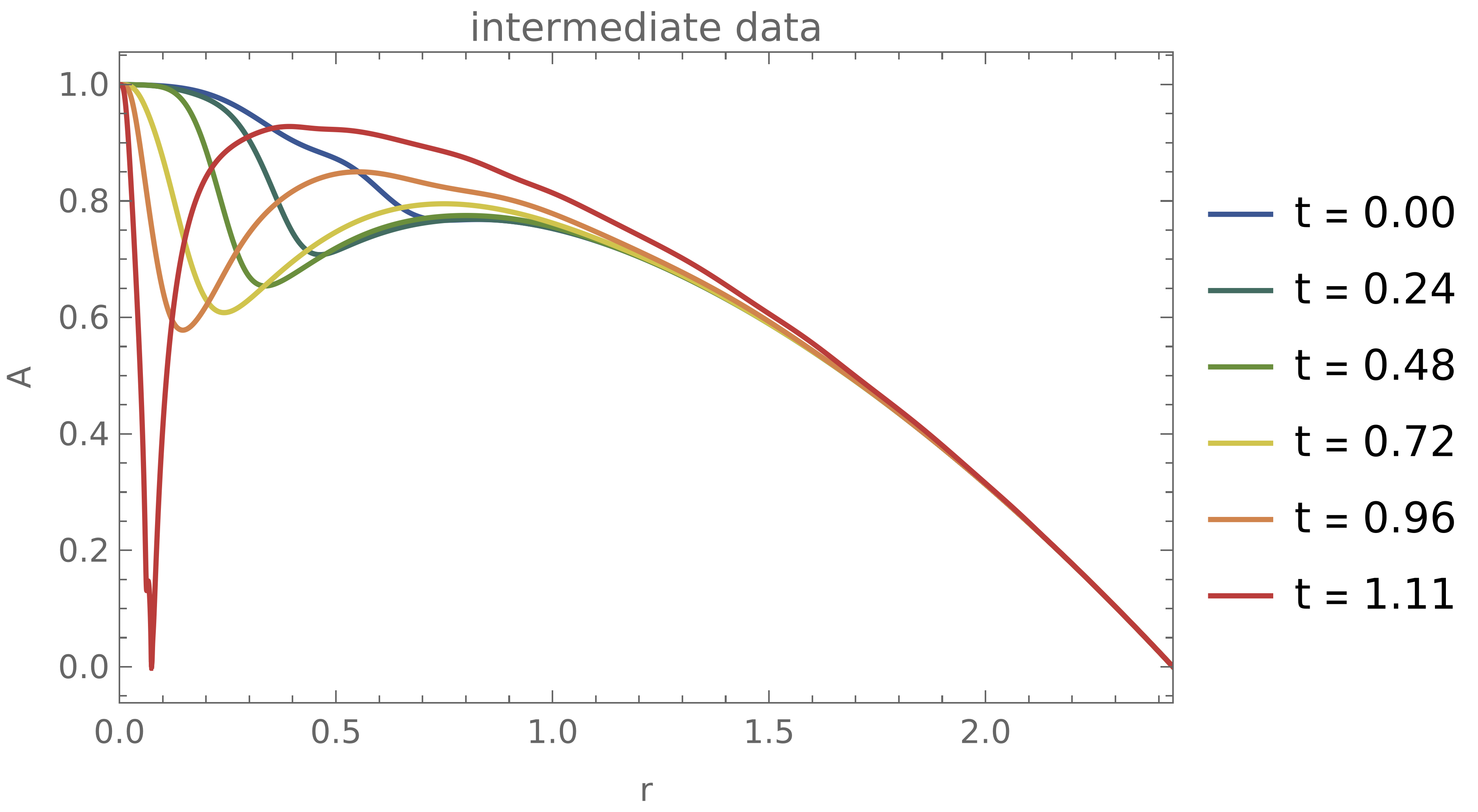}                     
    \caption{The function $A$ for intermediate initial data. The two zeros of $A$ correspond to the apparent $r_s=0.074$ and the cosmological $r_c=2.433$ horizons.}                                                                                          
    \label{fig:Ai}                                                               
    \end{minipage}

	\vspace{1cm}

    \begin{minipage}{\textwidth}
    \centering                                                                                                                                                             
    \includegraphics[scale=0.6]{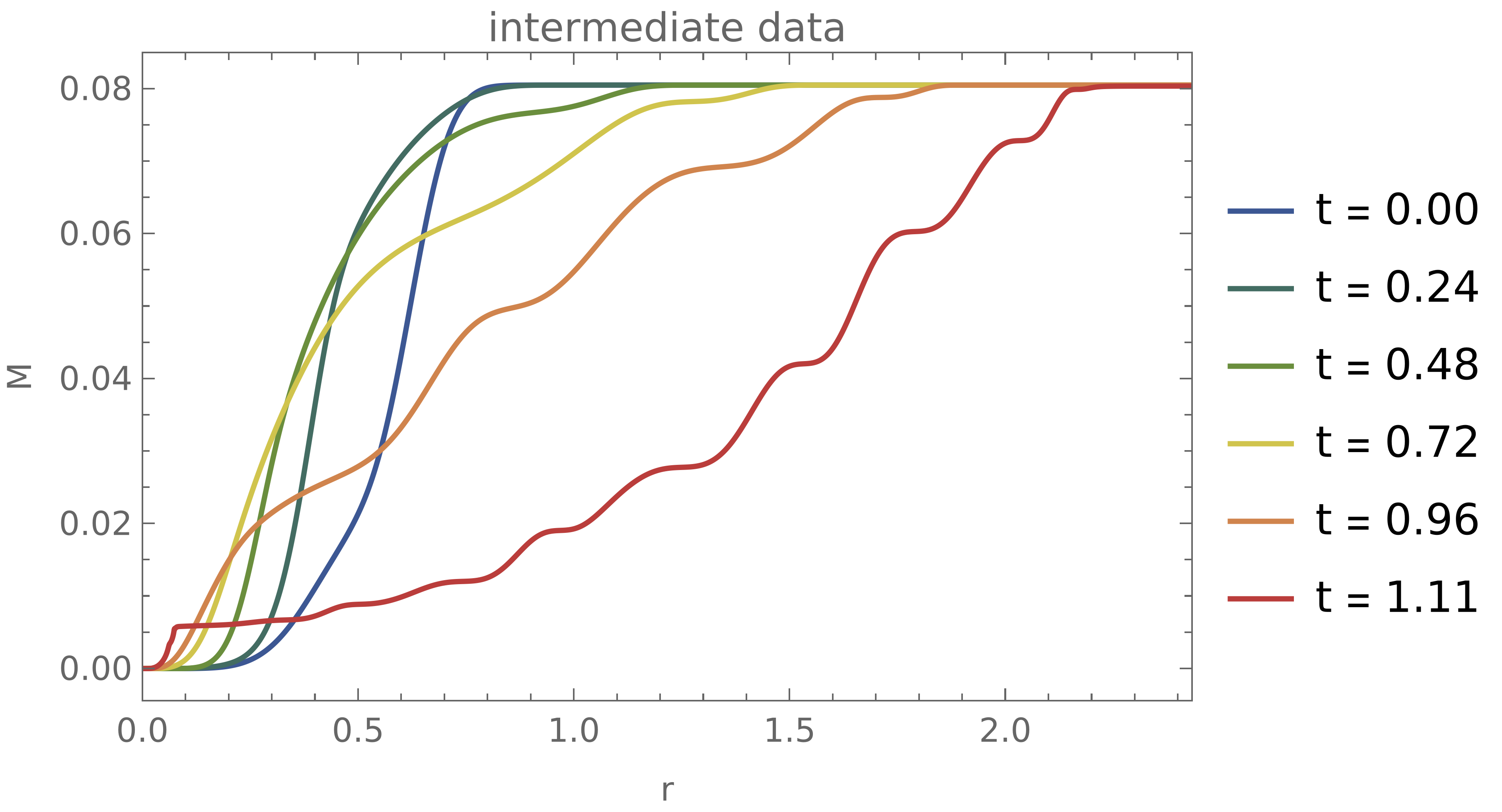}                     
    \caption{The Hawking mass $M$ for intermediate initial data.}                                                                                          
    \label{fig:Mi}                                                               
    \end{minipage}

\end{figure}

\begin{figure}[p]

    \centering                                                                                                                                                             
    \begin{minipage}[t][\textheight]{\textwidth}

    \centering                                                                                                                                                             
    \includegraphics[scale=0.6]{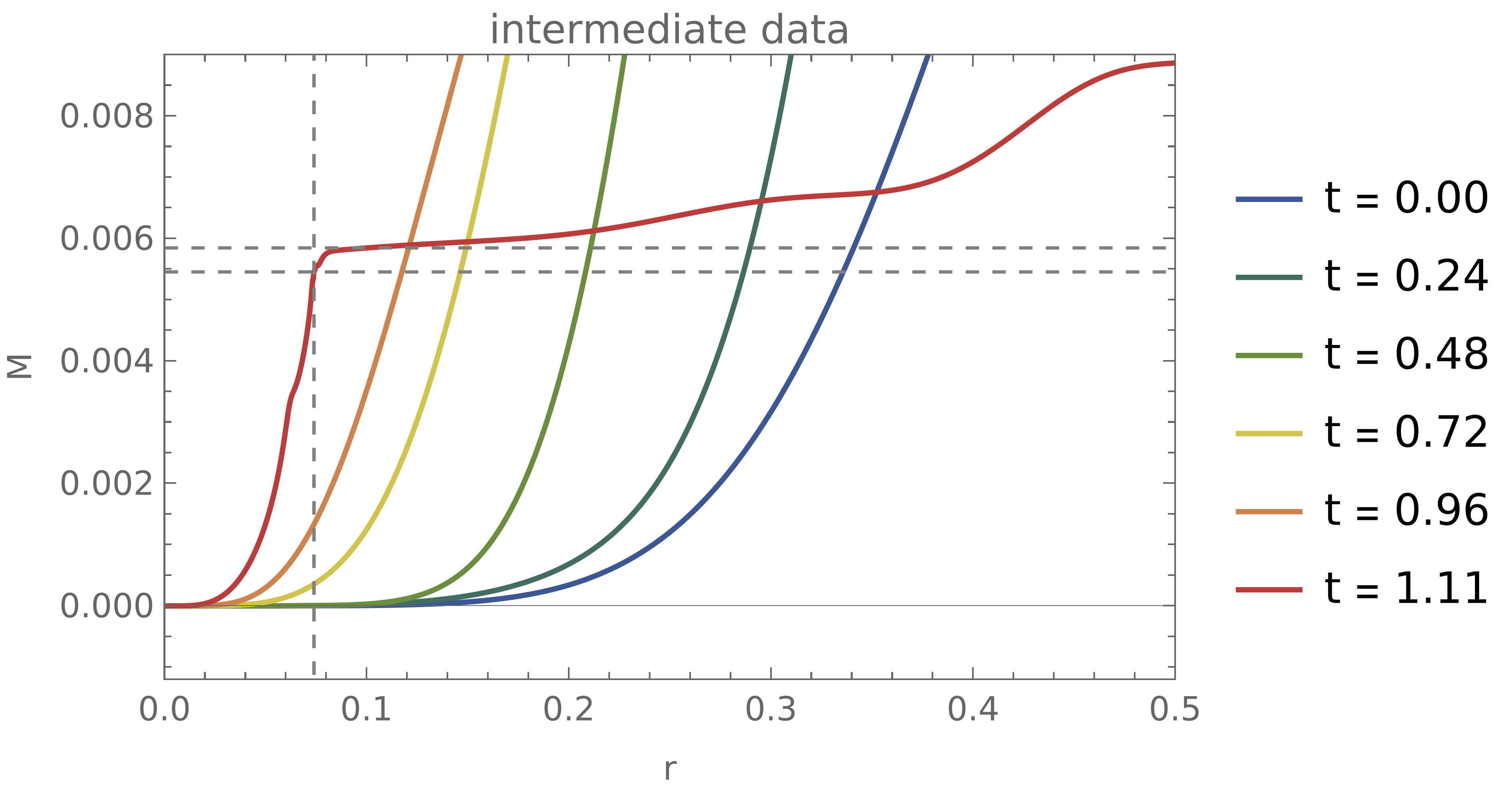}                     
	    \caption{The Hawking mass $M$ for intermediate initial data. Dashed lines indicate the apparent horizon mass, $M=0.0055$, the late-time plateau at $M=0.0058$, and the position of the apparent horizon.}
	    \label{fig:Mmi}                    

	    \vspace{1cm}

    \centering                                                                                                                                                             
    \includegraphics[scale=0.6]{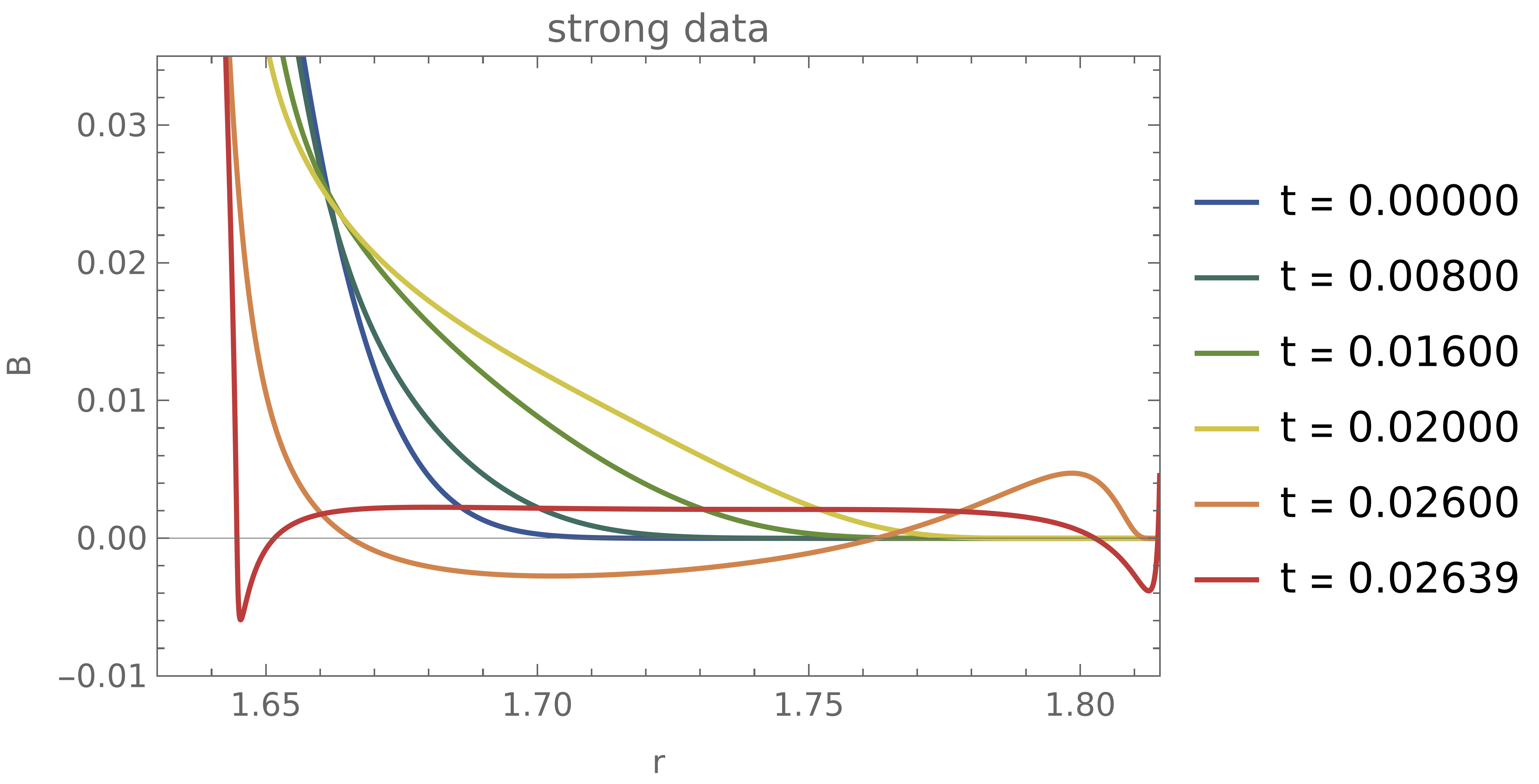}                     
    \caption{The squashing function $B$ for strong initial data.}                                                                                          
    \label{fig:Bs}                                                               

	    \vspace{1cm}

    \centering                                                                                                                                                             
    \includegraphics[scale=0.6]{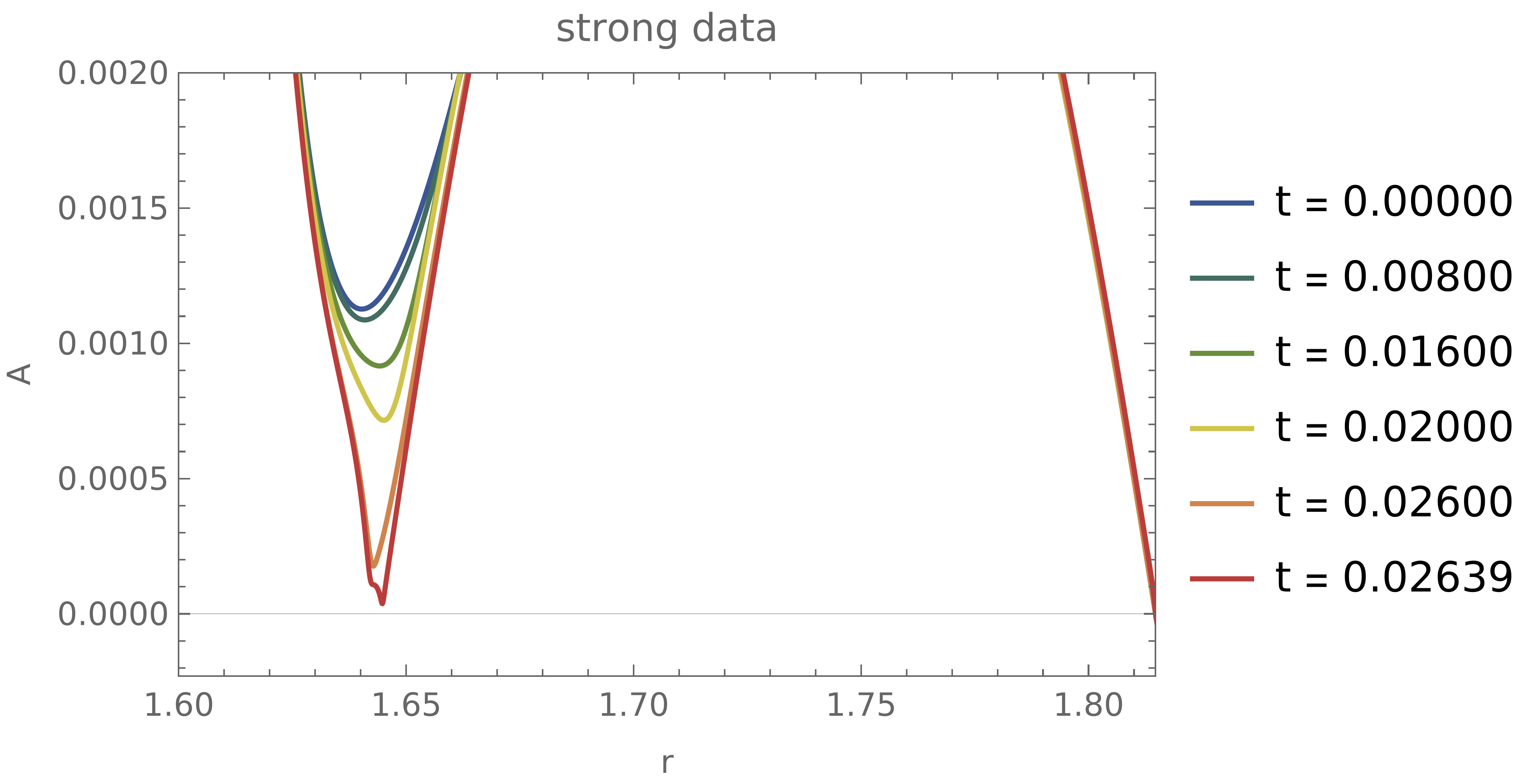}                     
    \caption{The function $A$ for strong initial data. The apparent horizon forms at $r_s=1.645$ where $A$ vanishes. The cosmological horizon is located at $r_c=1.815$ and corresponds to the second zero of $A$.}                                                                                          
    \label{fig:As}                                                               

    \end{minipage}

\end{figure}

Intermediate initial data with $b_0=1.091$, $r_0=0.05$, $s=0.5$, $v_0=-0.8$, $\Delta t=6\cdot 10^{-6}$, $\Lambda=1$, 
$r_{max}=2.44,
\Delta r=0.0000544655$
leads to collapse to a small black hole with the apparent horizon located at $r_s=0.074$. The cosmological horizon is at $r_c=2.433$. The evolution of $B$ is shown in Fig.\ \ref{fig:Bi}. The solution forms an apparent horizon with $M=0.0055$, and the Hawking mass $M$ forms a late-time plateau at slightly larger value  $M=0.0058$ as shown in Figs.\ \ref{fig:Mi}, \ref{fig:Mmi}. We expect this additional mass to be absorbed by the growing apparent horizon and the solution to settle down to an SdS black hole with $M=0.0058$. Since our code stops when a trapped surface forms, we cannot follow the subsequent evolution.

\begin{figure}[t]
\centering
    \includegraphics[scale=0.6]{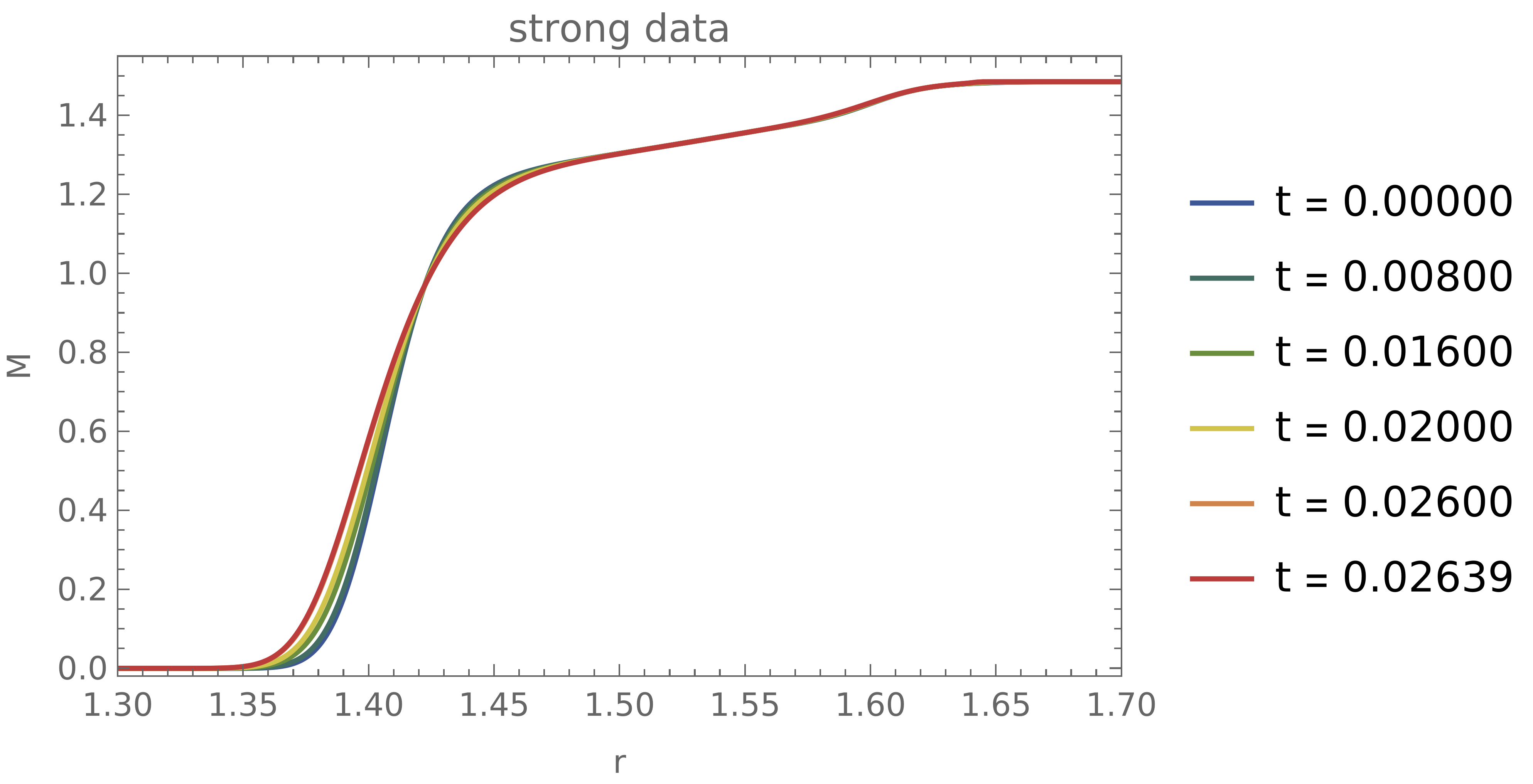}
    \caption{The Hawking mass $M$ for strong initial data.}
    \label{fig:Ms}
\end{figure}

\subsection{Strong data}

Strong initial data with $b_0=0.14$, $r_0=1.52$, $s=0.11$, $v_0=0.0$, $\Delta t=2\cdot 10^{-7}$, $\Lambda=1$, 
$
r_{max}=2.0,
\Delta r=0.0000446438$
lead to the collapse to a large black hole with the apparent horizon located at $r_s=1.645$. Since the cosmological horizon is located at $r_c=1.815$, the black hole is near--extremal with $r_s/r_c=0.906$. The final state is the SdS black hole with $M=1.485$, close to the extremal value of $M_{\mbox{max}}=1.5$. The evolution of $B$ is shown in Fig.\ \ref{fig:Bs}, and the function $A$ in Fig.\ \ref{fig:As}. The latter figure reveals how the apparent horizon forms at $r_s=1.645$ where $A$ vanishes. Initial data leading  to this near extremal evolution is already on the verge of forming trapped surface and its the Hawking mass $M$ is close to the extremal value at $t=0$, approaching this value at late times as shown in Fig.\ \ref{fig:Ms}. The coordinates are singular at both horizons and ill-defined in the limit $r_s\to r_c$, so the extremal black hole can not be reached in the numerical evolution, but the results suggest that it can be approached arbitrarily closely.


\section{Characteristic gluing}
\label{section_gluing}
The Kehle--Unger existence proof \cite{KU} of the gravitational collapse to extreme
RN metric
is based on the characteristic
gluing method developed in \cite{ACR}. A null cone in Minkowski  or Schwarzschild space--time, is glued to an extremal horizon in RN.
The resulting null cone is then extended to a four--dimensional region of a space--time using the results of
\cite{Luk}.  In this section we shall explain how the cosmological BCS ansatz and its numerical implementation which we have developed in
\S\ref{section_numerics} fit into this framework.

In what follows we shall consider two rectangular regions: $R_{dS}$ and $R_{SdS}$ isometric to subsets of
the de-Sitter space (dS) the Schwarzschild  de-Sitter (SdS) space respectively, both in static patch. 
Expressing the metrics in these regions in double---null coordinates (\ref{gEF}) we shall glue the outgoing dS null cone
in $R_{dS}$ to the black--hole horizon in $R_{SdS}$ both with
$u=0$ into one null cone $C_0$. This is achieved by making an ansatz for the $C^k$ function $B(u=0, v)$  in the gluing region $v\in [0, 1]$, where $v\leq 0$
in $R_{dS}$ and $v\geq 1$ in $R_{SdS}$, and such that $B(0, 0)=B(0, 1)=0$ as these values correspond to
dS and SdS spaces.

Once this is done the function $r(0, v)$ can be found from the constraint equation (\ref{emmm1})
\be
\label{raycha}
r_{vv}=-2r{B_v}^2
\ee
where we will use a freedom of redefining the outgoing null coordinate $v$  (this is  referred to as the lapse normalised gauge  in \cite{KU}) to set $F(0, v)=0$. The initial conditions for the ODE (\ref{raycha})
consist of $r(0, v)$ and $\p_v r(0, v)$ at either $v=0$ or $v=1$, and following the method of \cite{KU}
we shall choose $v=1$ which for us corresponds to the black--hole horizon $r_s$ of the SdS space. Once
$r(0, v)$ is determined and is continuous at $u=0$ together with its $v$--derivatives up to some fixed 
order $k$
we will be able to use the Einstein equations and their derivatives to find $u$--derivatives of $r, F , B$ at $u=0$ up to order $k$. This completes the gluing procedure. Now the characteristic 
data on the $u=0$ cone $C_0$ together with the characteristic data on the transverse null hypersurface $v=0$ in $R_{dS}$ can be used to construct a region of the space time with $u>0$ and $v>0$. Similarly the characteristic data on $C_0$ and the transverse null surface $v=1$ in $R_{SdS}$ can be used to construct a region of a space--time with $u<0$  and $v<1$. This is illustrated in Fig.\ \ref{fig:ChG}, and the details are given in the following subsections.
\subsection{Static solutions and their sphere data}
The Schwarzschild de--Sitter solution to (\ref{emmm1})--(\ref{emmm4}) is 
\[
B=0, \quad e^{2F}=1-\frac{m}{r^2}-\frac{1}{6}\Lambda r^2,
\]
where $r=r(u, v)$ is implicitly given by the relation
\be
\label{sdssol}
\exp\Big((v-u)\sqrt{1-\frac{2}{3}\Lambda m}\Big)=
\Big(\frac{r-r_s}{r+r_s}\Big)^{r_s} \Big(\frac{r_c +r}{r_c-r}\Big)^{r_c},
\ee
and black--hole and cosmological horizons $r_s, r_c$ are given by (\ref{rsrc}). There are three limiting cases
\begin{itemize}
\item Schwarzschild 
\be
\label{ssol}
e^{v-u}=e^{2r}\Big(\frac{r-\sqrt{m}}{r+\sqrt{m}}\Big)^{\sqrt{m}} \quad \mbox{if}\;\; \Lambda=0.
\ee
\item  De-Sitter 
\be
\label{dssol}
r=\sqrt{\frac{6}{\Lambda}}\tanh{\Big(\frac{v-u}{2}\sqrt{\frac{\Lambda}{6}}\Big)}\quad\mbox{if}\;\; m=0.
\ee
\item
Extreme Schwarzschild--de--Sitter, $r_c=r_s=\sqrt{3/\Lambda}$.
\begin{eqnarray}
\label{esdssol}
e^{2F}&=&-\frac{1}{{6\Lambda}r^2}\Big(\frac{3}{\Lambda}-r^2\Big)^2, \quad B=0, \quad\mbox{where}\nonumber\\
\frac{v-u}{2}&=&\frac{9}{{r_c}^4}\Big(r_c\mbox{arctanh}{(r/r_c)}+\frac{r}{(r/r_c)^2-1}\Big).
\end{eqnarray}
Note that $e^{2F}\leq 0$ as there is no static region in the extreme SdS case (and see
\cite{podolsky} for the discussion of the Penrose diagram of extreme SdS in 3+1 dimensions.)
\end{itemize}
In fact (\ref{sdssol}) gives rise to the whole {\it sphere data} at a 
sphere $S_0$ at $(u=0, v\in [v_1, 0]\cup[1, v_2])$ which consists of $r, F, B$ and their $(u, v)$ derivatives up to order $k$ ($k+1$ for $r(u, v)$) at $S_0$. The sphere data is lapse normalised at a cone $u=0$ if $F$ and all its derivatives vanish at this cone. In what follows we shall only use the gauge independent part of
this sphere data, which consists of the signs of derivatives of $r$ with respect to $u$ and $v$.
On the Schwarzschild-de Sitter black hole horizon   $r(u=0, v)=r_s$ we get $r_v=0$ because
$r$ does not change with $v$. On the other hand $r_u$ measures the deviation in the direction transverse to the horizon, so is non--zero, and negative as the incoming null rays are converging.
\subsection{Gluing} Define rectangular regions $R_{dS}$ and $R_{SdS}$ of the de--Sitter and  Schwarzschild de--Sitter spaces respectively in by specifying the ranges of their respective double--null coordinates in static patches by
\[
R_{dS}=[0, u_2]\times [v_1, 0], \quad R_{SdS}=[u_1, 0]\times [1, v_2]
\]
where $u_1<0<u_2$ and $v_1<0<1<v_2$. The edge $[v_1, 0]$ of $R_{dS}$ is a null cone in de--Sitter space,  and the edge $[1, v_2]$ in $R_{SdS}$ is the black--hole horizon.
We restrict these regions to their null cones $u=0$ and glue these
cones to create an outgoing null cone $C_0$. The gluing will preserve the radial symmetry of the BCS ansatz,
so each point on $C_0$ is a topological $3$--sphere, and the points on
$R_{dS}\cap C_0$ and $R_{SdS}\cap C_0$ are round $3$--spheres which we shall call the de--Sitter spheres and Schwarzschild de--Sitter spheres respectively. The characteristic data has therefore been specified on the part of $C_0$ given by 
$\{u=0, [v_1,0]\cup [1, v_2]\}$ which consists of two disjoined intervals. The data in the region $u=0, v\in [0, 1]$ is specified by making an ansatz
 \be
\label{ansatsB}
B(u=0, v)=\sum_{j=1}^k  \alpha_j b_j(v)
\ee
for the squashing factor  in this region subject
to the continuity conditions at $0$ and $1$ which force the vanishing of $B$ and its derivatives
up to a specified order (which depends on the regularity of gluing we want to achieve)  at these 
$3$--spheres. Here $b_j$ are bump functions (different anzatses have different forms of $b_j$, see \cite{KU} and \cite{RS}), and the constants $\alpha_j$ need to be determined by vanishing of $u$--derivatives of $B(u, v)$  at $u=0, v=1$.

We shall exploit the coordinate freedom (these coordinate transformations leave the Misner-Sharp mass invariant) $v\rightarrow \tilde{v}(v)$ and 
$u\rightarrow \tilde{u}(u)$ to set
\[
F(0, v)=0, \quad (\p_u)^i F(0, 0)=0, \quad i=1, \dots, k.
\]
To determine $r, F$ and their $u$--derivatives, as well as the $u$ derivatives of $B$  on $C_0$ we turn to the Einstein equations (\ref{emmm1}--\ref {emmm4}) and (\ref{lasteeq}). We first solve the constraint equation
(\ref{emmm1}) 
backwards using the values of $r$ and $r_v$ at the SdS sphere $(u=0, v=1)$. These values can be 
read--off
from (\ref{sdssol}), so that the ODE we must solve is
\[
r_{vv}=-2r{B_v}^2  , \quad r(u=0, v=1)=r_s, \quad r_v(u=0, v=1)=0.
\]
The uniqueness theorem for second order ODEs then gives $r(u=0, v)$ on $C_0$. We will now solve
the ODEs resulting from the remaining Einstein equations  and their derivatives {\it forward} in $v$
propagating from $v=0$ to $v=1$.

First  turn to the wave equation (\ref{emmm3}) which becomes
\[
\p_v (r_{u}) =\frac{\Lambda r}{6}-\frac{2}{r}r_ur_v
+\frac{1}{6r}(e^{-8B}-4e^{-2B}).
\]
This is a first order ODE for $r_u(0, v)$. The initial 
condition $r_u(0, 0)$ follows from the vanishing of the Misner-Sharp mass  
at the dS three--sphere 
$(u=0, v=0)$ which gives
\be
\label{ruat0}
r_u(0, 0)=\frac{\Lambda r(0, 0)^2-6}{24 r_v(0, 0)}.
\ee
We have so far determined $r$ and $r_u$ at $C_0$.
Another wave equation (\ref{emmm4}) now becomes
a 1st order ODE for $B_u(0, v)$ which we solve with the initial data $B_u(0,  0)=0$. We must make sure
that the resulting $B_u(1, v)=0$ to match the sphere data at the SdS horizon. This places a condition on the parameters in the ansatz (\ref{ansatsB}), which we fix by a shooting procedure.

The final wave equation
(\ref{lasteeq}) is a 1st order ODE for $F_u(0, v)$. 
The function $F$ is equal to $0$ at $C_0$ - this is our gauge choice. The function $F_u$ is determined by solving
(\ref{lasteeq})
with an initial condition $F_u(0, 0)=0$. 
Proceeding in an analogous way, and using the $u$ and $v$ derivatives of the Einstein equation we can
determine the $u$--derivatives of $r, F, B$ at $C_0$.

\begin{figure}[t]
\begin{center}
\includegraphics[scale=0.3,angle=0]{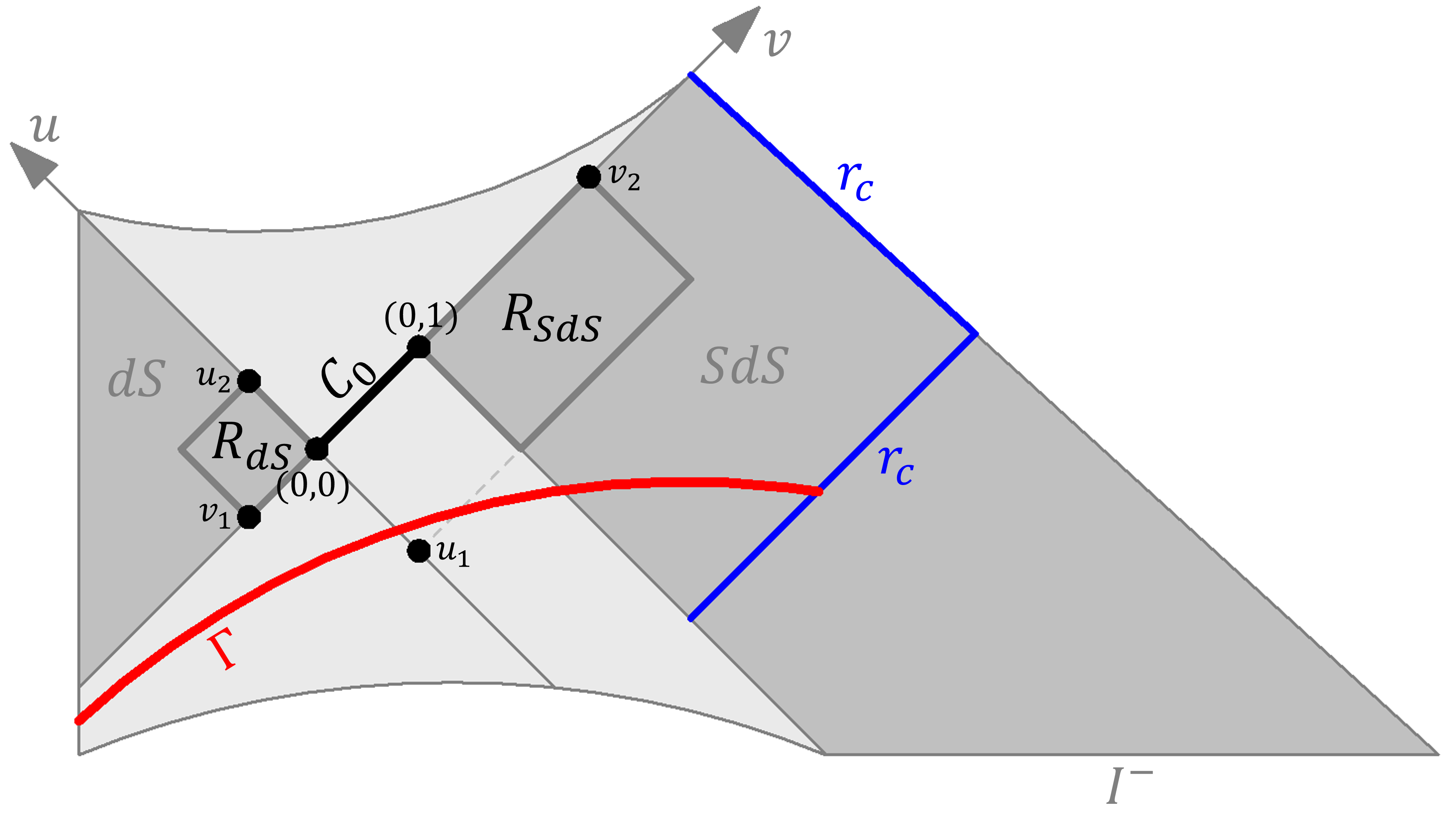}
\begin{center}
\caption{Characteristic gluing and the space--time extension. A de--Sitter null cone in $R_{dS}$ and
the black hole horizon in  $R_{SdS}$
are glued along the cone $C_0$ given by $u=0$ by making an ansatz for  the squashing factor $B$ at $u=0, v\in [0, 1]$ and using the ODE constraints to find the rest of the sphere data. This data on $C_0$ is then used, together with the data on two transverse null edges of $R_{dS}$ and $R_{SdS}$ to extend the space--time away from $C_0$. Extending backwards allows to read off the Cauchy data on the space--like
surface $\Gamma$ used in \S\ref{section_numerics} to solve the Einstein equations numerically. In our setup
this Cauchy data is only specified up to the cosmological horizon.}
\label{fig:ChG}
\end{center}
\end{center}
\end{figure}

The data on $C_0$ has been determined. We also have the transverse null surface data at
the SdS edge $(v=1, u\in[u_1, 0])$. A theorem of Luk \cite{Luk} allows extension to $u<0$, and once this is found we can read off the Cauchy data at a space--like surface $t=0$. This is the data  we use in our numerical implementation in \S\ref{section_numerics} to construct the space--time. This initial data is therefore determined, by a combination of the characteristic gluing construction followed by the extension theorem.
Conversely, the numerical evolution of the  Cauchy data from \S\ref{section_numerics} determines  
$B(u=0, v)$ needed for the characteristic gluing, as well as $B(u>0, v)$ which in the framework of \cite{Luk} would be determined by the data on $C_0$ and the transverse null surface $(v=0, u\in [0, u_2])$.
This is illustrated in Fig.\ \ref{fig:ChG}.

\subsection{$C^1$ and $C^2$ gluing}
For a fixed value of $\Lambda$ which defines the de-Sitter patch in the gluing procedure we consider
the ansatz  (\ref{ansatsB}) of the form
\be
\label{restricteda}
B=v^2(1-v)^2(1+\alpha v^2).
\ee
The parameter $\alpha=\alpha(m)$ depends on the mass of the SdS patch which in turn determines the black--hole radius $r_s$ given by (\ref{rsrc}) at $(u=0, v=1)$. For a given $m$ we specified $\alpha$ by a shooting argument requiring
\[
(\p_u B)(u=0, v=1)=0.
\]
We find that $\alpha$ varies very slowly with $m$. For $\Lambda=0.1$ a few values are
\[
(m, \alpha)=(1, -3.2115), \quad(5, -3.2103), \quad  (10, -3.2081), \quad (15, -3.2017),
\]
with the final pair corresponding to the extremal SdS parameters. For all subextremal configurations
we find that $r>0$ and $r_u<0$ on $v\in[0, 1]$ and $u=0$. These conditions are necessary for the metric to remain non--degenerate and without an anti--trapped surface. 
A modification involving two constants $(\alpha_1, \alpha_2)$
\be
\label{Broydentwoconstants}
B=v^2(1-v)^2(1+\alpha_1 v+\alpha_2 v^2)
\ee
combined with the Broyden method  leads to a  $C^2$ solution with $B_u$ and $B_{uu}$ 
vanishing at $(0, 1)$. In the extremal case $m=15$ we find $\alpha_1=-2.89671$, $\alpha_2=1.63196$. 
(See Fig.\ \ref{fig:C2rru})
\subsection{$C^0$ gluing to the Nariai solution} 
\begin{figure}[htbp]
\begin{center}
\includegraphics[width=0.48\textwidth,angle=0]{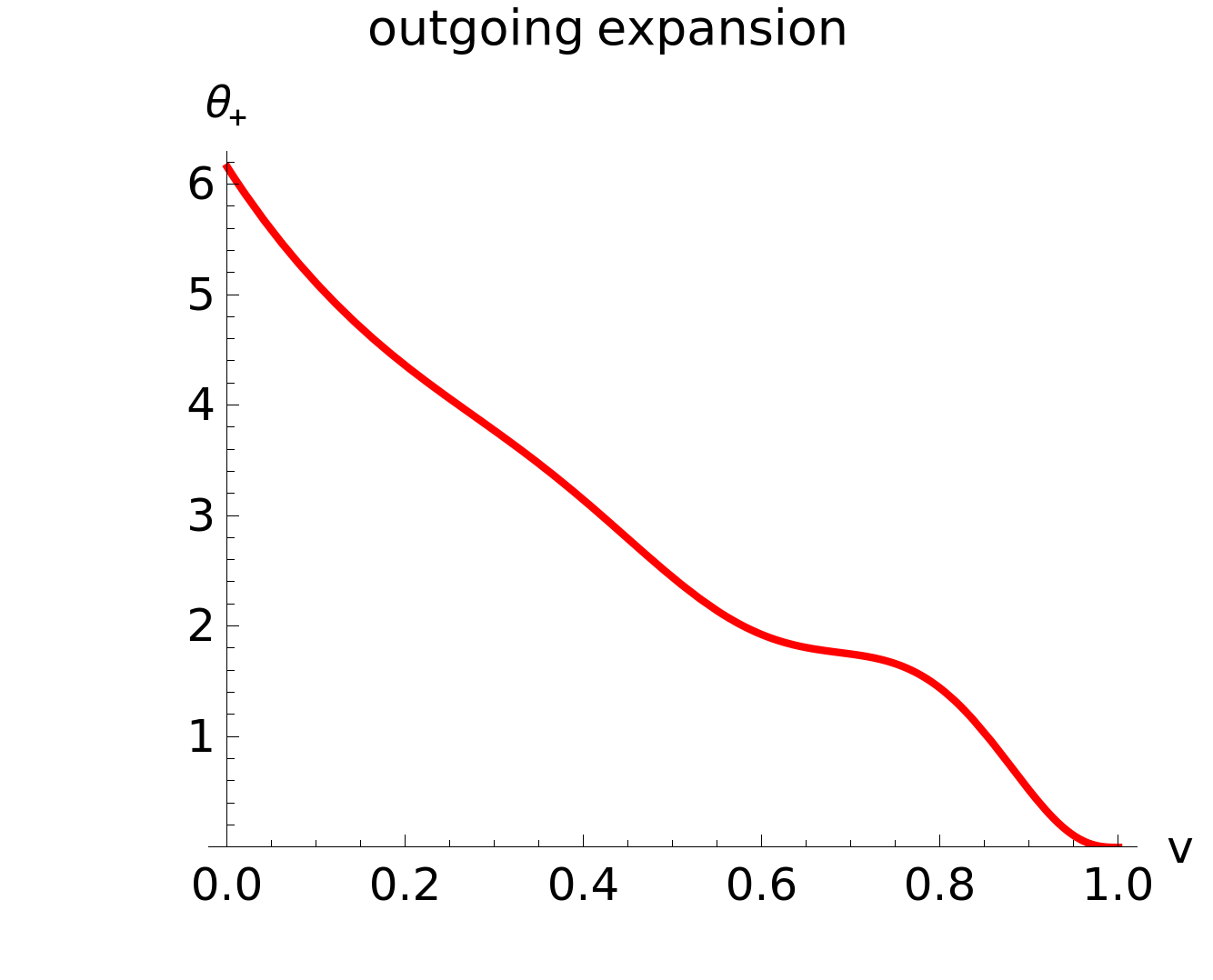}\hfill\includegraphics[width=0.48\textwidth,angle=0]{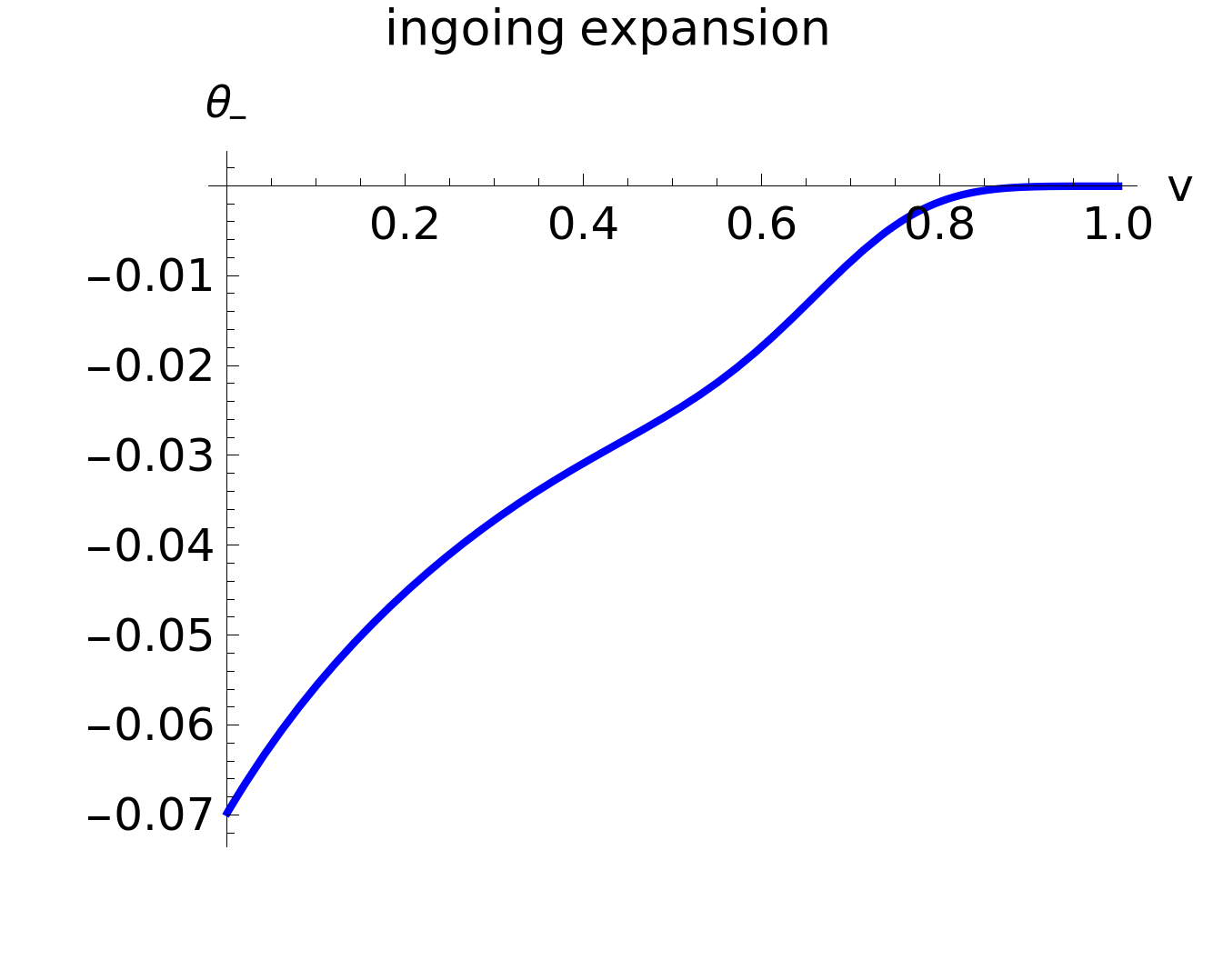}
\begin{center}
\caption{ Outgoing and ingoing expansions $\theta_+=3r^{-1} r_v, \theta_-=3r^{-1} r_u$
on the gluing surface $u=0$ with Nariai condition at $u=0, v=1$, and $\Lambda=0.1, m=15$.}
\label{fig:theta}
\end{center}
\end{center}
\end{figure}

The extremal Schwarzschild-de Sitter space time does not admit a bifurcation sphere where $r_u$ and $r_v$ simultaneously vanish. There is a near--horizon limit $dS^2\times S^3$ known as the Nariai space--time,
which is regular and has $r_u=r_v=0$ at each point.
While one does not expect this Nariai space--time to form in a gravitational collapse, it is instructive 
to see whether $r_u(0, 1)=0$ arises from characteristic gluing. We consider the same ansatz (\ref{restricteda}) but rather than insisting on matching $B_u$ at the extreme SdS horizon we settle for $B$ being just continuous, but instead force
$
r_u(0, 1)=0
$
(see Fig.\ \ref{fig:theta}.)
Implementing the bisection method on $\alpha$ we find, for $\Lambda=0.1, m=15$,
$
 \alpha=-14.1179, \quad r_u(0, 1)\sim 3.72\times10^{-9}\,.
$
\subsection{Profile functions along the gluing cone}
\begin{figure}[htbp]
\begin{center}
\includegraphics[width=0.48\textwidth,angle=0]{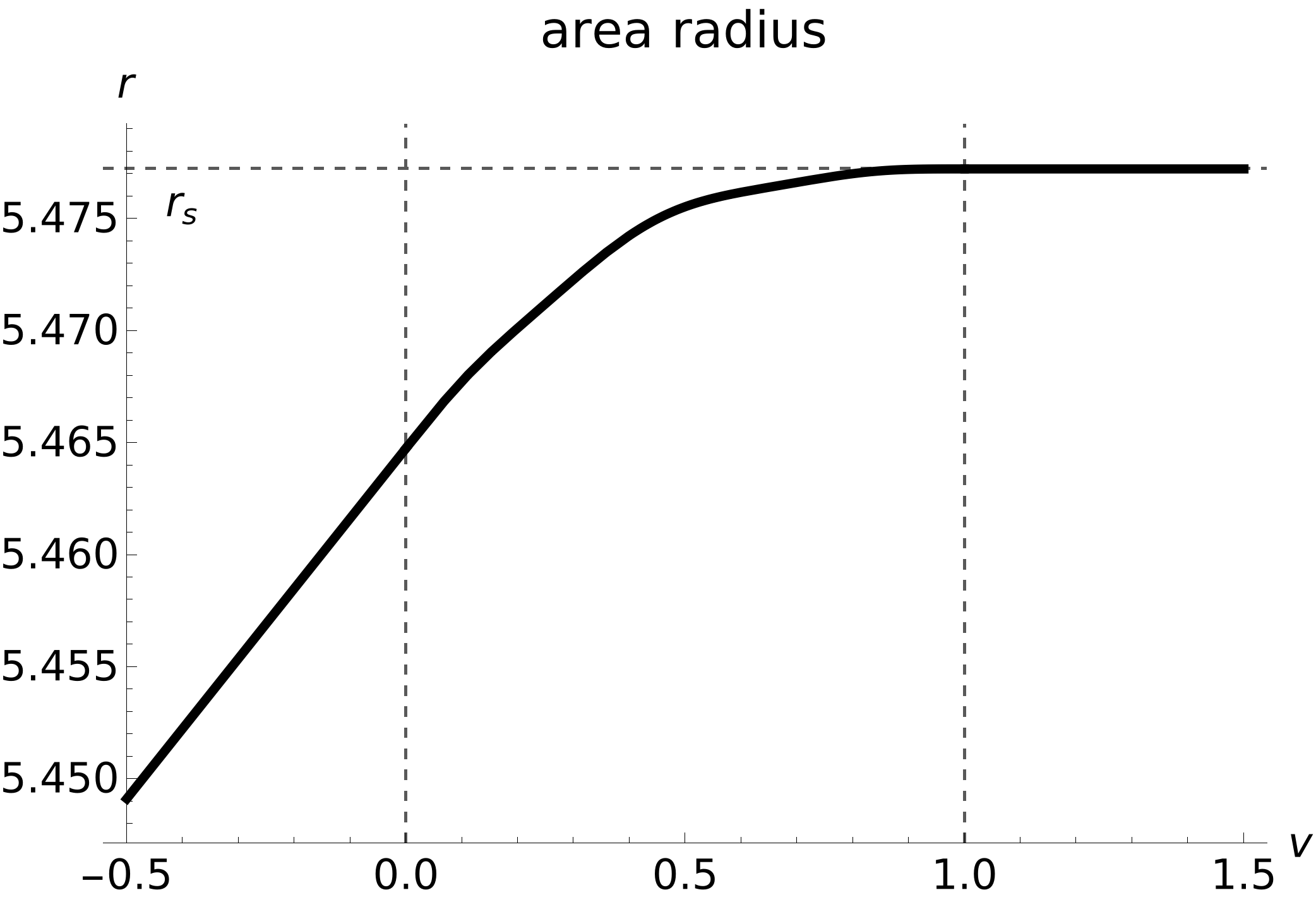}\hfill\includegraphics[width=0.48\textwidth,angle=0]{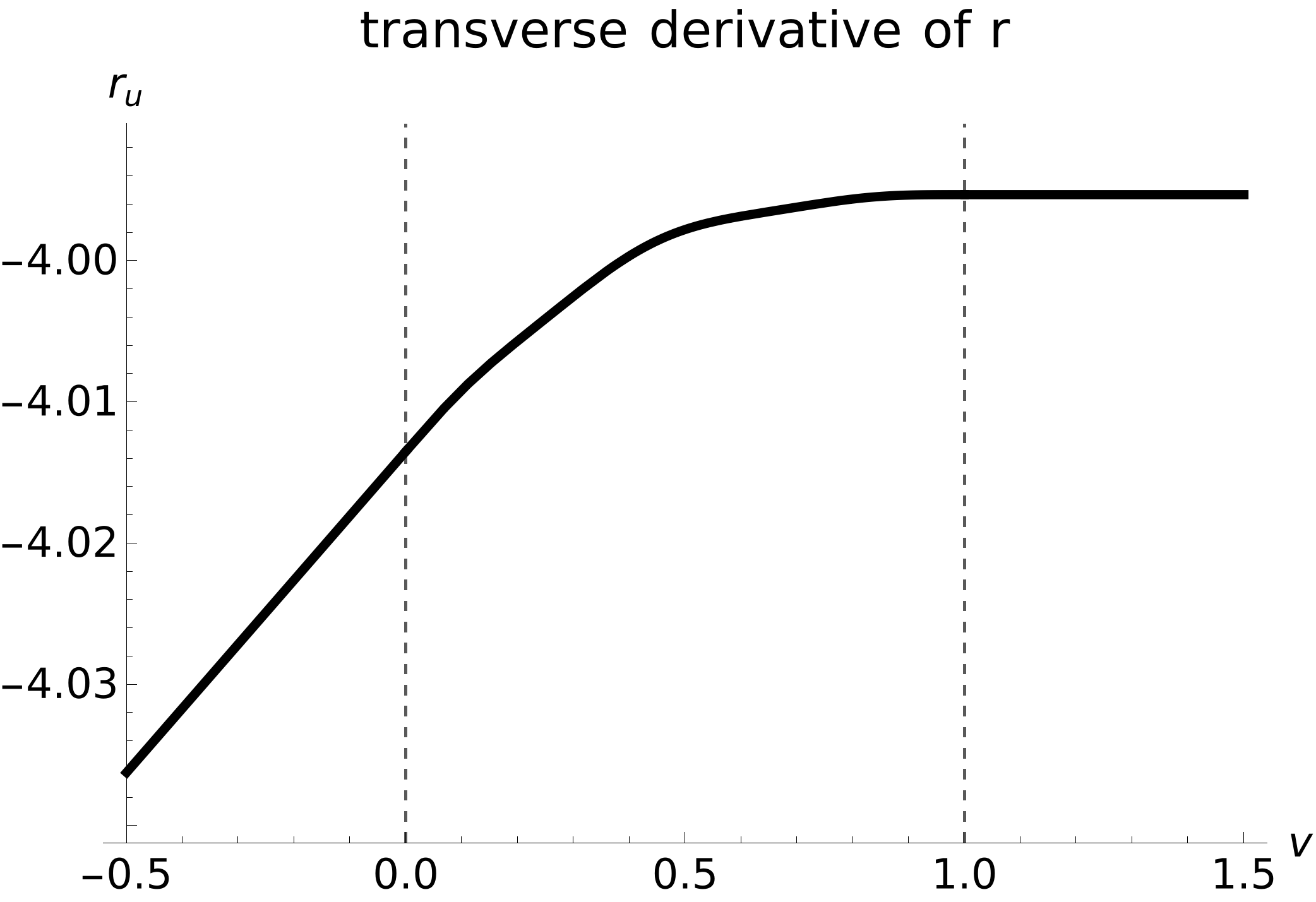}
\begin{center}
\caption{The area radius $r(0,v)$ (left) and its transverse derivative $r_u(0,v)$ (right)
on the null cone $u=0$ in the vicinity of the gluing region $v\in[0,1]$ (vertical dashed
lines), for $\Lambda=0.1$, $m=15$ and the ansatz $B(0,v)=v^2(1-v)^2(1+\alpha_1\,v+\alpha_2\,v^2)$ with
$\alpha_1=-2.89671$, $\alpha_2=1.63196$. The radius attains $r_s=\sqrt{30}$ (horizontal dashed line)
at $v=1$ and remains constant along the horizon, where $r_u$ is constant as well.}
\label{fig:C2rru}
\end{center}
\end{center}
\end{figure}

\begin{figure}[htbp]
\begin{center}
\includegraphics[width=0.48\textwidth,angle=0]{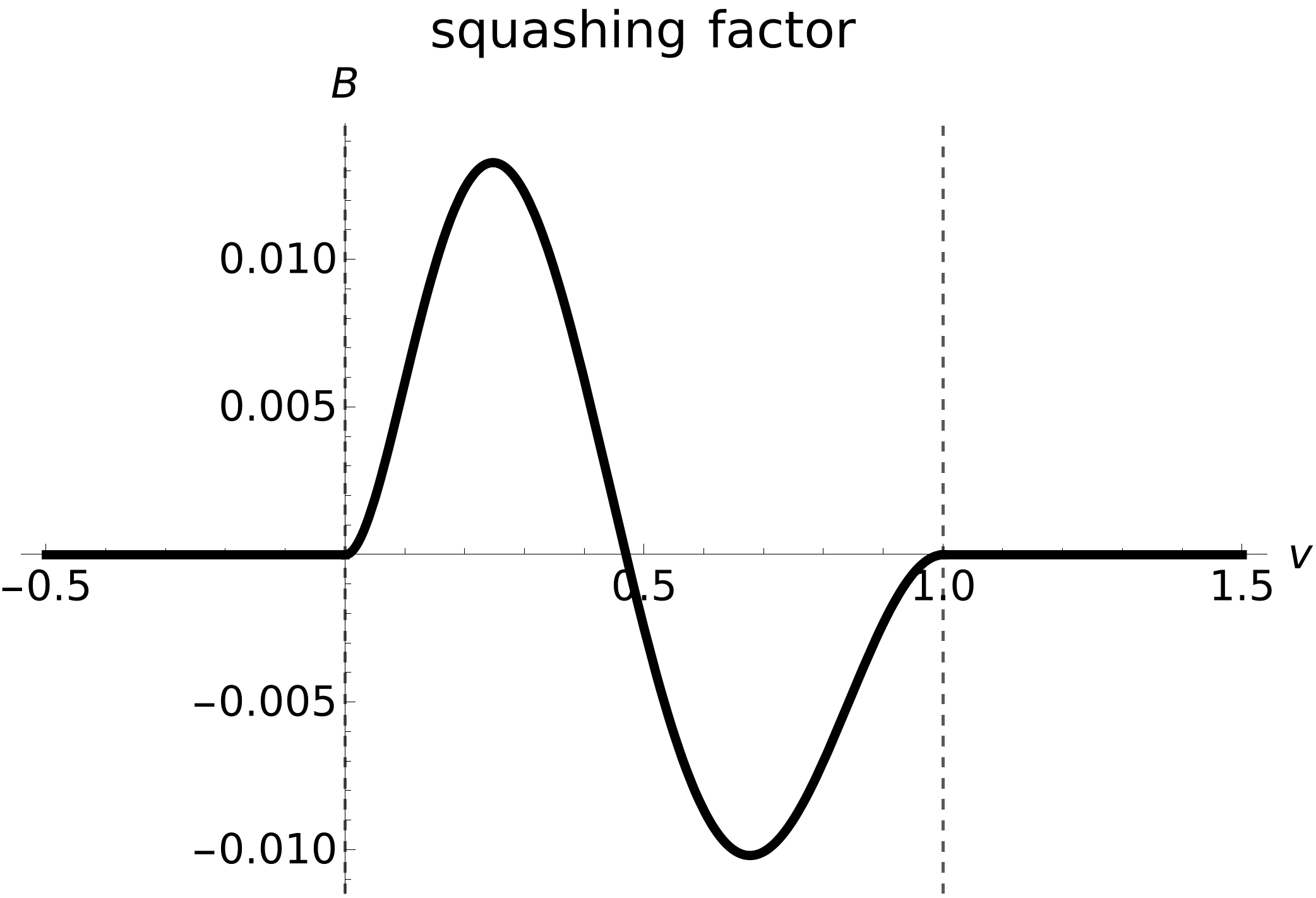}\hfill\includegraphics[width=0.48\textwidth,angle=0]{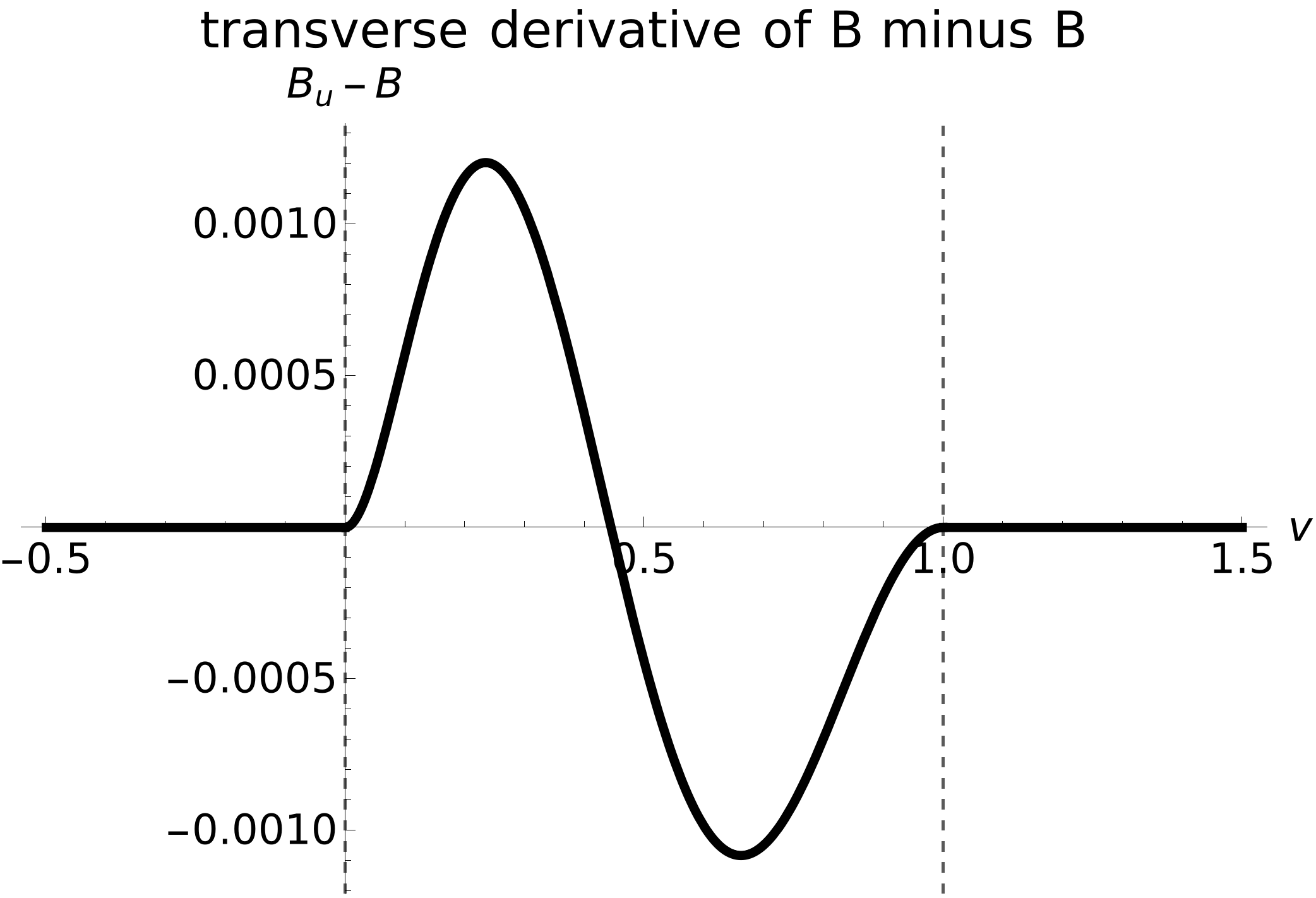}
\begin{center}
\caption{The squashing factor $B(0,v)$ (left) and the difference
$B_u(0,v)-B(0,v)$ (right) for the data of Figure \ref{fig:C2rru}. The
transverse derivative $B_u$ closely follows $B$ on the gluing region --- the
difference is an order of magnitude smaller than either function
($\max|B_u-B|\simeq 1.2\times 10^{-3}$, while
$\max|B_u|\simeq 1.4\times 10^{-2}$) --- so we plot $B_u-B$ instead of $B_u$.
Both $B$ and $B_u$ vanish for $v\leq 0$ and $v\geq 1$; in particular, this
choice of $c$, $d$ satisfies the matching condition $B_u(0,1)=0$.}
\label{fig:C2BBu}
\end{center}
\end{center}
\end{figure}

\begin{figure}[htbp]
\begin{center}
\includegraphics[width=0.48\textwidth,angle=0]{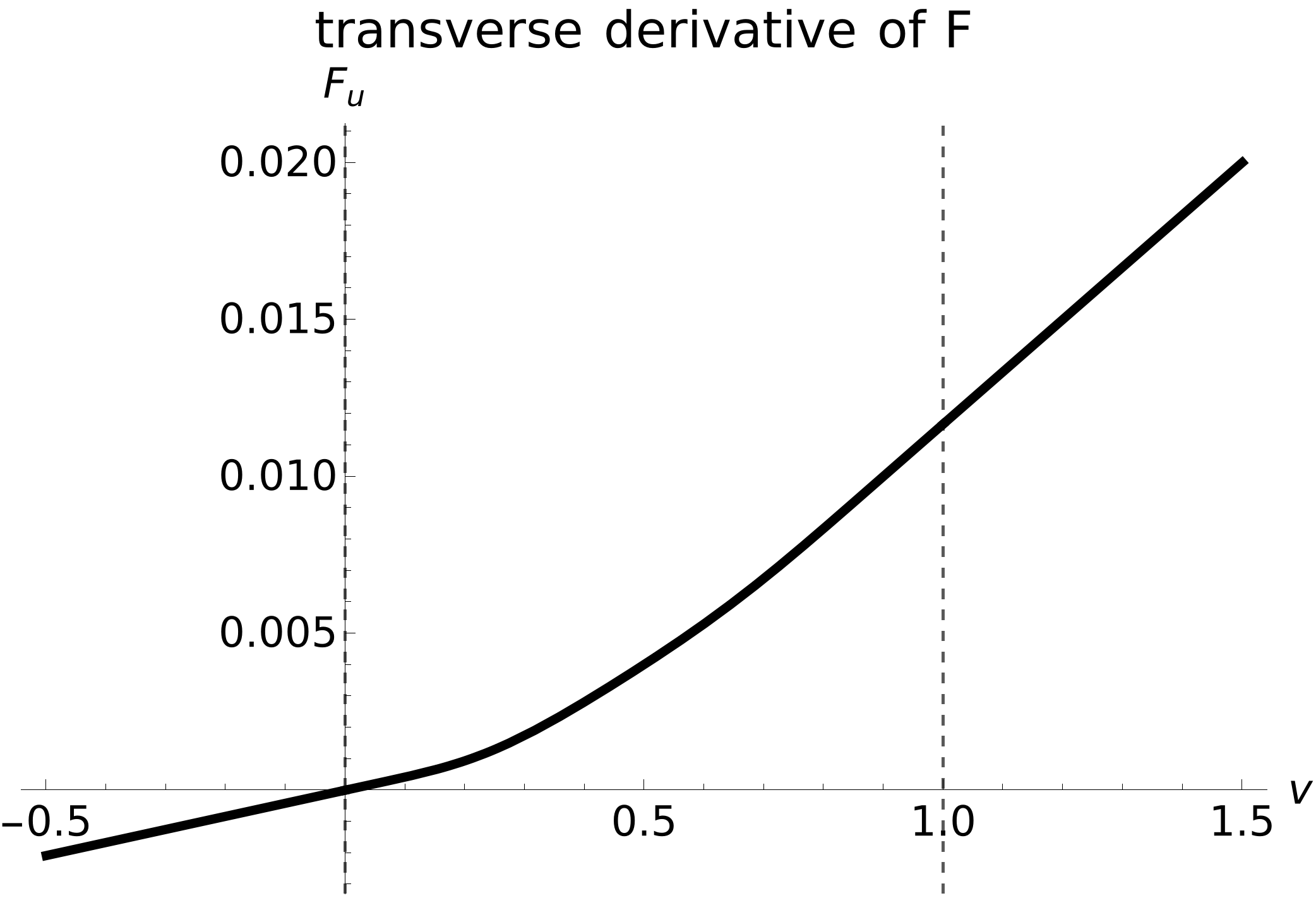}\hfill\includegraphics[width=0.48\textwidth,angle=0]{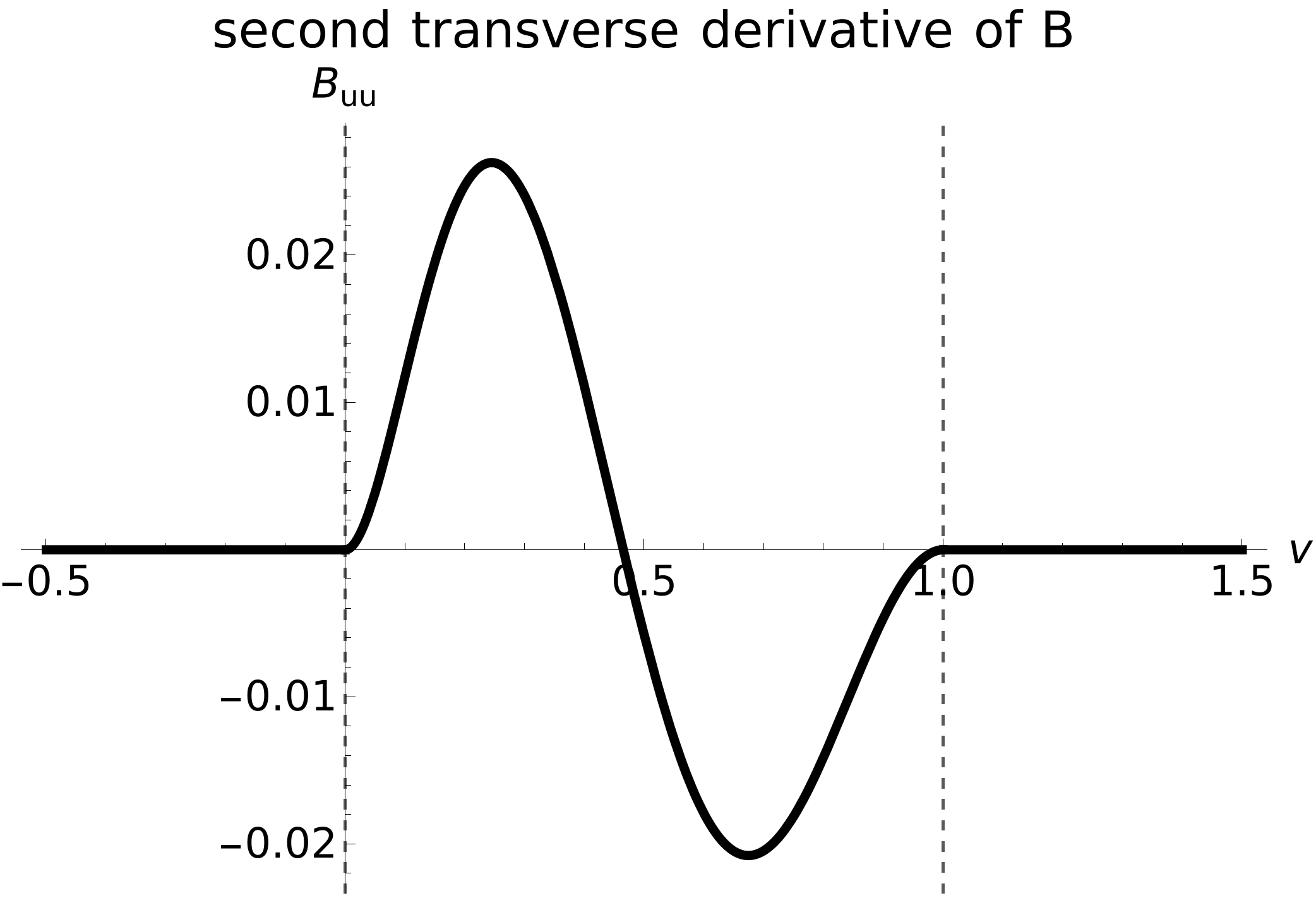}
\begin{center}
\caption{The transverse derivative $F_u(0,v)$ (left) and the second transverse
derivative $B_{uu}(0,v)$ (right) for the data of Figure \ref{fig:C2rru}
($F(0,v)=0$ is the gauge choice). The slope of $F_u$ equals $\Lambda/24$ in the
de--Sitter region and $\Lambda/6$ along the horizon. The function $B_{uu}$
vanishes for $v\leq 0$ and is constant for $v\geq 1$; its value $B_{uu}(0,1)=0$
is the second matching condition, so this choice of $c$, $d$ glues the
squashing factor in a $C^2$ manner.}
\label{fig:C2Fu}
\end{center}
\end{center}
\end{figure}

In what follows we restrict  to the extreme case $r_s=r_c=\sqrt{3/\Lambda}$, i.e.\ $\Lambda m=\frac{3}{2}$. The constraint (\ref{emmm1}) in the gauge $F(0,v)=0$, $F_u(0,0)=0$ for the de--Sitter case ($u=0$, $v<0$) reads $r_{vv}=0$. Therefore, $r(0,v)=r(0,0)+r_v(0,0)\,v$. Similarly, the Schwarzschild--de Sitter case ($u=0$, $v>1$) gives in our gauge $r_{vv}=0$, so that $r(0,v)=r_s$ for $v>1$. The vanishing of the Misner-Sharp mass $M_\Lambda$ in the de--Sitter region extends (\ref{ruat0}) to the whole de--Sitter region, so that $$r_u(0, v)=\frac{\Lambda r(0, v)^2-6}{24 r_v(0, 0)}\,,$$
there. For the Schwarzschild--de Sitter region, the equation (\ref{emmm3}) gives $r_{uv}(0,v)=\frac{\Lambda r_s^2-3}{6 r_s}=0$, hence $r_u(0,v)=r_u(0,1)$ for $v>1$. In both regions $B(0, v)=0$, as this is a part of the de--Sitter and Schwarzschild--de Sitter sphere data, hence also $B_v(0, v)=0$ 
there. In the de--Sitter region (\ref{emmm4}) reduces to $B_{uv}=-\frac{3r_v}{2r}B_u$, which together with the initial condition $B_u(0,0)=0$ gives $B_u(0,v)=0$. In the Schwarzschild--de Sitter region, where additionally $r_v=0$, Eq.\ (\ref{emmm4}) reduces to $B_{uv}=0$, hence $B_u(0, v)=B_u(0, 1)$ for $v>1$. The constant $B_u(0, 1)$ is determined by the forward integration of (\ref{emmm4}) through the gluing region $v\in[0, 1]$. The matching to the extreme Schwarzschild--de Sitter sphere data requires $B_u(0, 1)=0$: this is the $C^1$ matching condition enforced by the shooting procedure, which however cannot be combined with $r_u(0,1)=0$ within the one--parameter ansatz. Eq.\ (\ref{lasteeq}) reduces $F_{uv}=\frac{3}{r^2}(\frac{1}{4}+r_u r_v)-\frac{\Lambda}{12}$. For the de--Sitter case, $F_{uv}=\frac{\Lambda}{24}$, hence $F_u(0,v)=F_u(0,0)+\frac{\Lambda}{24}\,v=\frac{\Lambda}{24}\,v$. For the Schwarzschild--de Sitter case, $F_{uv}=\frac{3}{4r_s^2}-\frac{\Lambda}{12}=\frac{\Lambda}{6}$, hence $F_u(0,v)=F_u(0,1)+\frac{\Lambda}{6}\,(v-1)$ for $v>1$. These results are summarized in Table \ref{tab:C0data}. The sphere data in the vicinity of the gluing region is plotted in Figures \ref{fig:C2rru}--\ref{fig:C2Fu}.
\begin{table}[h!]
\centering
\renewcommand{\arraystretch}{2.2}
\begin{tabular}{l|c|c}
 & de--Sitter & extreme Schwarzschild--de Sitter \\
\hline
$(u,v)$ & $u=0,\quad v\in\big[-\frac{r(0,0)}{r_v(0,0)},\,0\big]$ & $u=0,\quad v\in[1,\,v_2]$ \\
\hline
$r$ & $r(0,0)+r_v(0,0)\,v$ & $r_s=\sqrt{3/\Lambda}$ \\
$B$ & $0$ & $0$ \\
$F$ & $0$ & $0$ \\
$r_u$ & $\dfrac{\Lambda\, r(0,v)^2-6}{24\, r_v(0,0)}$ & $r_u(0,1)$ \\
$B_u$ & $0$ & $B_u(0,1)$ \\
$F_u$ & $\dfrac{\Lambda}{24}\,v$ & $F_u(0,1)+\dfrac{\Lambda}{6}\,(v-1)$ \\
\end{tabular}
	\caption{The sphere data on the null cone $u=0$ in the gauge $F(0,v)=0$, $F_u(0,0)=0$. The constants $r(0,0)$ and $r_v(0,0)>0$ follow from the
backward integration of the constraint (\ref{emmm1}) in the gluing region
	$v\in[0,1]$ with $r(0,1)=r_s$, $r_v(0,1)=0$. The constant $F_u(0,1)$ follows from the forward integration of Eq.\ (\ref{lasteeq}) with $F_u(0,0)=0$. The centre $r=0$ corresponds
to the vertex of the de--Sitter cone at $v=-r(0,0)/r_v(0,0)$. In the
extreme case the surface gravity vanishes, so $r_u$ and $B_u$ are
constant for $v>1$. }
\label{tab:C0data}
\end{table}

\FloatBarrier
\section{Outlook}
The presence of a non--zero cosmological constant $\Lambda$ has a quantitative effect on gravitational collapse, slowing it down if $\Lambda>0$ (see  \cite{shapiro, joshi}) and enhancing it if $\Lambda<0$.
The latter case has been investigated within the BCS ansatz \cite{BR}.  In asymptotically  
$AdS_5$ the waves bounce back from the $AdS$ boundary, so
there is no dispersion  and a black hole 
forms for any size of initial data. This is consistent with the non--linear instability of $AdS$. In our paper
we have focused on $\Lambda>0$, where both the dispersion and the collapse can occur depending on the size of the initial data and, unlike the $\Lambda=0$ case, the size of the black hole horizon is limited by the cosmological horizon. We have shown numerically that large initial data can result in a near extremal collapse,
but offered no proofs that an extremal collapse does take place. A  mathematical work in this
direction is under way \cite{monica}. Other numerical works on near extremal collapse we are aware of 
\cite{harvey, Gelles} focus on charged black holes, where the collapse is driven by a scalar field.

 We have not investigated the other extreme, where the initial data is small
and disperses back to the de-Sitter space. It would be interesting to do so in the context of critical cosmological collapse, as most of the research in this field  has been done in the absence of cosmological constant \cite{review}.

\section*{Acknowledgments}
The numerical implementation used in this work builds on an older code due to Zbisław Tabor, adapted by Tadeusz Chmaj and Sebastian J. Szybka for the BCS model \cite{Szybka}. We used \textit{Claude Code} to resolve  technical issues and to implement tools for visualizing the results. Some calculations were performed using Wolfram Mathematica and the xAct package \cite{xAct}. We are grateful to Piotr Bizo\'n,  Mihalis Dafermos,
Maxime Gadioux,
Mónica Tapia del Moral and
Claude Warnick 
for discussion and correspondence. We also thank
the anonymous referee for comments that led to improvements.
\section*{Data availability statement}
The C++ code is not publicly available
but may be shared by the authors for verification purposes upon a reasonable
request. This research has not involved studies on human subjects, human data or tissue, or animals. Authors are not aware of any competing interests.

\section*{Appendix A: Linear perturbation theory and quasi--normal modes}
\appendix
\renewcommand{\theequation}{A.\arabic{equation}}
\label{section_QNM}
The ringdown phase of settling into an Schwarzschild-de Sitter black hole in the gravitational collapse is dominated by the least damped quasi--normal mode (QNM) of the linearised Einstein equations. In this Appendix we shall find an analytical formula for the QNM frequencies in the near--extremal Schwarzschild--de--Sitter case.

\begin{figure}[htbp]
\begin{center}
\includegraphics[scale=0.6,angle=0]{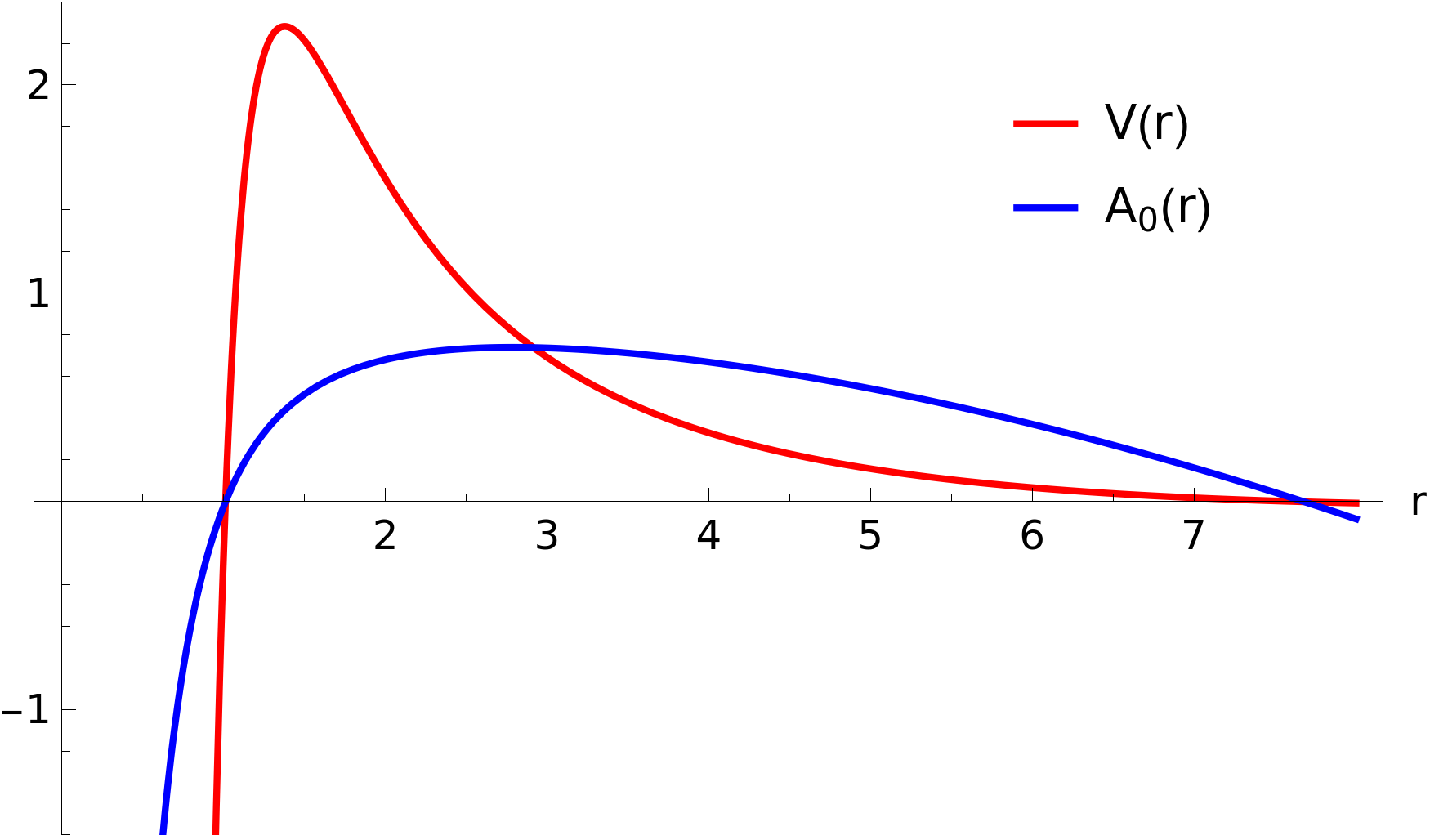}
\caption{$V(r)$ in red and $A_0(r)$ in blue with $m=1, \Lambda=0.1$. The zeroes $r_s$ and $r_c$ of $A_0$ correspond to the black hole and the cosmological horizons.}
\label{fig:VA}
\end{center}
\end{figure}

Linearising (\ref{em3}) with $B(r, t)=\beta(r, t)$ around the static solution (\ref{static}) gives
\be
\label{betalin}
\ddot{\beta}-\frac{1}{r^3}A_0(r^3A_0\beta')'+
\frac{8}{r^2}A_0\beta=0.
\ee
Setting
\be
\label{xformula}
x=\int {A_0}(r)^{-1}dr, \quad \beta=e^{-ikt} r^{-3/2} u(x)
\ee
turns (\ref{betalin}) into the Schr\"odinger equation
\be
\label{schrodinger}
-\frac{d^2 u}{dx^2}+V(r(x))u=k^2u,
\ee
where
\begin{eqnarray}
\label{potential}
V&=&\frac{A_0}{4r^2}(6r{A_0}'+3A_0+32)\\
&=&\frac{1}{4r^2}\Big(1-\frac{m}{r^2}-\frac{1}{6}\Lambda r^2\Big)\Big(\frac{9m}{r^2}
-\frac{5\Lambda}{2}r^2+35\Big).\nonumber
\end{eqnarray}
The potential $V(r)$ and the attractor $A_0(r)$ given by (\ref{static})
are plotted in Fig.\ \ref{fig:VA}.
The asymptotic of convergence to Schwarzschild-de Sitter in the gravitational collapse is dominated by the least damped quasi--normal mode of  (\ref{schrodinger}) with complex frequency
$k=k_R-ik_I$ with $k_I>0$ in the units of ${r_s}^{-1}$, where $r_s$ is the location of the black--hole horizon. It can be found numerically, or using either the WKB or the P\"oschl--Teller approximation.  
\begin{itemize}
\item The WKB method \cite{SW} gives
\[
k^2=V(x_0)-i\sqrt{-2V_{xx}(x_0)}(n+\frac{1}{2}),
\]
where $V(x_0)$ is the local maximum of $V$ attained at $x_0$, and 
$V_{xx}(x_0)=V''(r(x_0)) ({A_0}')^2$. 
\item The P\"oschl--Teller method relies on approximating $V$ by the potential $V_{PT}$ for which the QNMs are known 
analytically
\be
\label{PSapprox}
V_{PT}=\frac{V_0}{\mbox{cosh}^2((x-x_0)/b)},\quad
k=\frac{1}{b}\Big(\sqrt{V_0b^2-\frac{1}{4}}-
(n+\frac{1}{2})i\Big)
\ee
where $V_0=V(x_0)$ and $b$ is found by demanding that $V$ and $V_{PT}$ have the same second derivatives at $x_0$. 
\end{itemize}
In the  near--extremal Schwarzschild de--Sitter case  an  analytical formula for QNM frequencies
can be obtained as follows:
Let $r_s$ and $r_c$ be given by (\ref{rsrc}). Assuming  $(r_c-r_s)/r_s\ll 1$ 
and expanding $A_0$ to the second order in the range $r\in [r_s, r_c]$ yields
\begin{eqnarray}
\label{A0form}
A_0&=&\frac{\Lambda}{6r^2}(r_c-r)(r-r_s)(r_c+r)(r+r_s)\nonumber\\
&&\sim\frac{2\kappa_s}{r_c-r_s}(r_c-r)(r-r_s).
\end{eqnarray}
In this derivation we assumed that $r+r_s, r+r_c$
and $r^{-2}$ vary slowly on the interval 
$[r_s, r_c]$. Then $A_0=(r_c-r)(r-r_s)\alpha(r)$ for some $\alpha(r)$.
Differentiating this relation and expanding near $r=r_s$ we find $\alpha(r)\sim 2\kappa_s/(r_c-r_s)$
where $\kappa_s\equiv \frac{1}{2}{{({A_0})}_r}|_{r=r_s}$ is the surface gravity at the black hole horizon.
Substituting (\ref{A0form}) into (\ref{xformula}) leads to the tortoise coordinate $x$, and an inversion formula
\[
r\sim \frac{1}{2}(r_c+r_s)+\frac{1}{2}(r_c-r_s)\tanh{(\kappa_s x)}, \quad
A_0\sim \frac{\kappa_s(r_c-r_s)}{2}\frac{1}{\cosh^2{(\kappa_s x)}}.
\]
We now turn to the potential (\ref{potential}). The 
first two terms in the second bracket lead to contributions of order $(r_c-r_s)^2$ which can be neglected. The last term is  
$8\Big(\frac{r_c+r_s}{2}\Big)^{-2}$, so that finally $V$ can be approximated by the P\"oschl--Teller potential
\[
V\sim \frac{V_0}{\cosh^{2}{(\kappa_s x)}}, \quad
\mbox{where}\quad
\kappa_s\sim \frac{r_c-r_s}{{r_s}^2}, \;\;
V_0=4{\kappa_s}^2.
\]
Using (\ref{PSapprox}) we now find the lowest QNM to be 
\be
\label{final_QNM}
k=\frac{1}{2}\kappa_s(\sqrt{15}-i), \quad\mbox{where}\quad\kappa_s \sim \sqrt{\frac{\Lambda}{3}}\sqrt{1-\frac{2}{3}\Lambda m}.
\ee
A similar calculation of QNMs, but in 3+1 dimensions, has been performed in \cite{Cardoso}. The analysis in 4+1 dimensions appears to be simpler because of the {double quadratic} form
(\ref{static}) of $A_0(r)$.

\FloatBarrier
\section*{Appendix B: Crank-Nicolson scheme and convergence}
\appendix
\renewcommand{\theequation}{B.\arabic{equation}}
\label{section_Crank}
The spatial discretisation is second--order accurate (a centred/finite--volume
discretisation of the flux and trapezoidal constraint
integration, both $O(\Delta r^{2})$).  The functions $B$ and $P$ are advanced by
a Crank--Nicolson rule for fixed metric coefficients, but the full
constraint--evolution coupling is in general first order in time because the metric
coefficients are lagged (if the metric coefficients do not vary rapidly, then the code becomes effectively second--order accurate in time).

We have tested the convergence directly (similarly as in \cite{PHA}) for the three
initial--data sets used in the paper (weak, intermediate and strong). We used two
independent diagnostics on \emph{five} nested grids
$N=801,1601,3201,6401,12801$ over the fixed domain $r\in[10^{-6},2.5]$
(refinement ratio two, so coarse nodes coincide with fine ones). For each data
set the Courant ratio $\Delta t/\Delta r$ ($0.1$ for weak and intermediate, $0.05$ for strong initial data) is held fixed across the five grids
(so a single exponent governs the combined space--time error), the output times
are matched, and the window is chosen before any trapped surface forms. The
binary is compiled without \code{-ffast-math} so the arithmetic does not
contaminate the order measurement.

The two diagnostics are: \textbf{(i)} the $\Delta r$--weighted $L_2$ norm of the
\emph{independent residual} of the redundant Einstein equation
\begin{equation}
  E(r)\;=\;\dot A + 4\,r\,A\,\dot B\,B',
\end{equation}
which is not imposed during the evolution and therefore measures the truncation
error directly; and \textbf{(ii)} the three--grid Cauchy self--convergence of the
squashing function $B$ $$Q=\|B_h-B_{h/2}\|_2/\|B_{h/2}-B_{h/4}\|_2\,,$$ with
order $\log_2 Q$ (the same $\Delta r$--weighted $L_2$ norm as in (i)).
Both diagnostics are evaluated on the interior nodes only: the first and the
last grid node are excluded, as their update is governed by the boundary
treatment (the regularised centre $r=10^{-6}$ and the outer boundary) rather
than by the interior scheme.

\paragraph{Result.}
The weak (dispersing) and strong (near--extremal) data converge effectively at second order,
while the intermediate data converges at first order. The independent residual at
the final time $t^*$ falls as

\begin{figure}[t!]
\centering
\includegraphics[width=0.70\textwidth]{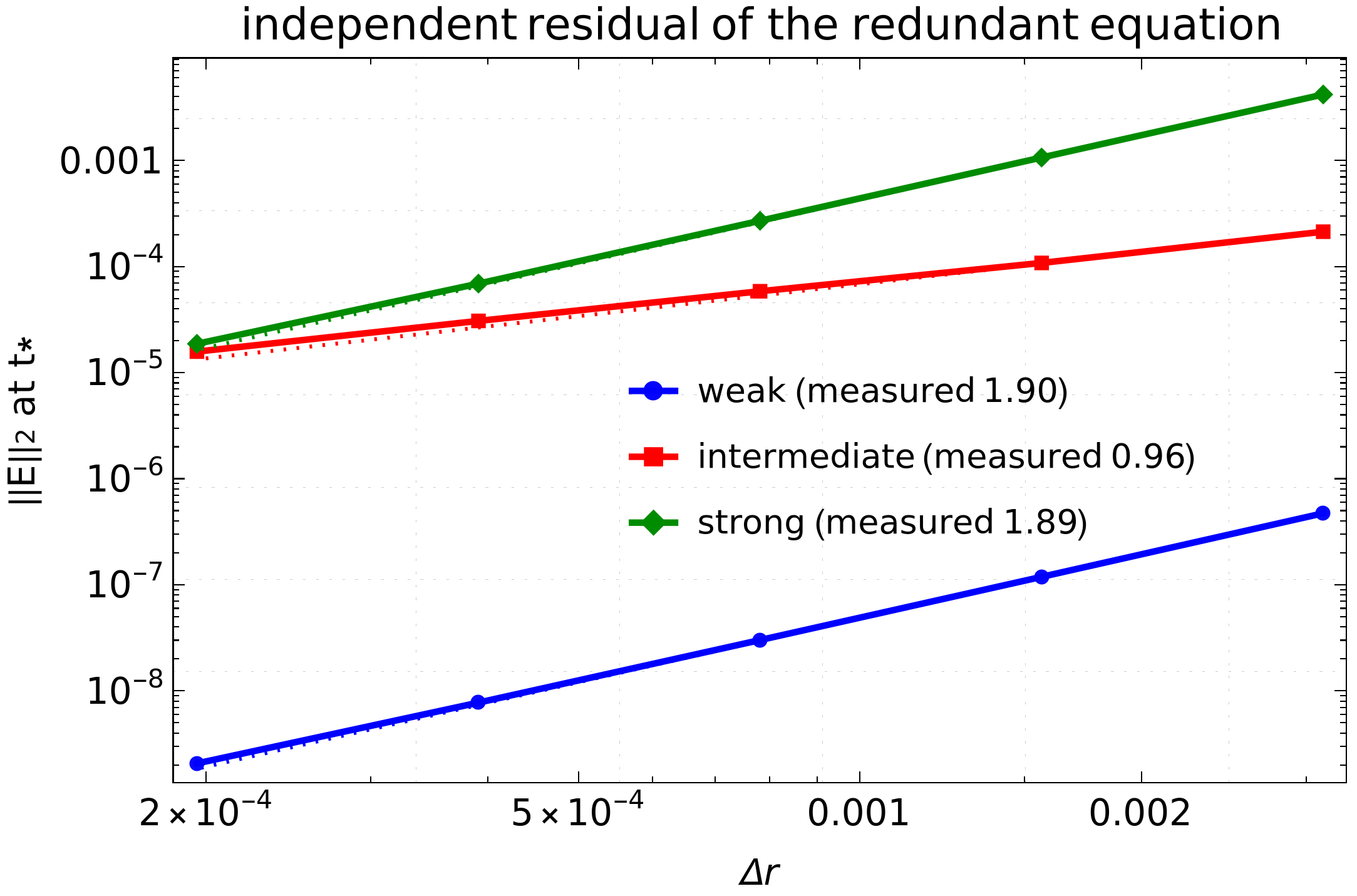}
\caption{Independent residual $\|E\|_2=\|\dot A+4rA\dot B\,B'\|_2$ of the
redundant Einstein equation at the matched final time, versus the grid spacing
$\Delta r$, for the three data sets on five nested grids ($N=801\ldots12801$).
Dotted lines are reference slopes ($2$ for weak and strong, $1$ for
intermediate). The weak and strong data converge at second order, the
intermediate data at first order.}
\label{fig:convres}
\end{figure}

\begin{figure}[t!]
\centering
\includegraphics[width=0.96\textwidth]{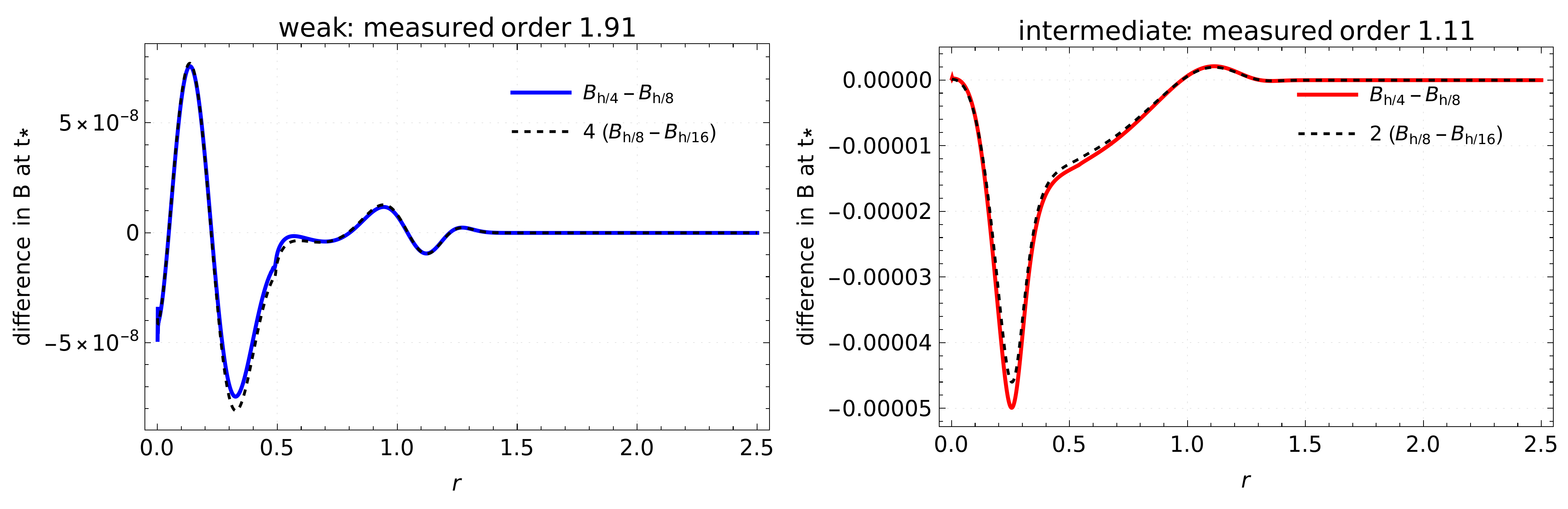}
\caption{Three--grid scaled--difference self--convergence of $B$ at $t^*$ on the
finest grid triple. For the weak data (left) $B_{h/4}-B_{h/8}$ overlaps
$4\,(B_{h/8}-B_{h/16})$, the signature of second order; for the intermediate data
(right) it overlaps $2\,(B_{h/8}-B_{h/16})$, the signature of first order.}
\label{fig:convself}
\end{figure}

\begin{center}
\begin{tabular}{lccccc}
 & \multicolumn{5}{c}{$\|E\|_2$ on the five grids} \\
\hline
N & $801$ & $1601$ & $3201$ & $6401$ & $12801$\\
\hline
weak ($t^*{=}0.5$)         & $4.7\!\times\!10^{-7}$ & $1.2\!\times\!10^{-7}$ & $3.0\!\times\!10^{-8}$ & $7.8\!\times\!10^{-9}$ & $2.1\!\times\!10^{-9}$\\
intermediate ($t^*{=}0.5$) & $2.1\!\times\!10^{-4}$ & $1.1\!\times\!10^{-4}$ & $5.9\!\times\!10^{-5}$ & $3.1\!\times\!10^{-5}$ & $1.6\!\times\!10^{-5}$\\
strong ($t^*{=}0.0175$)     & $4.2\!\times\!10^{-3}$ & $1.1\!\times\!10^{-3}$ & $2.7\!\times\!10^{-4}$ & $6.9\!\times\!10^{-5}$ & $1.9\!\times\!10^{-5}$\\
\hline
\end{tabular}
\end{center}

\noindent i.e.\ $\|E\|_2$ drops by $\approx4$ per halving for the weak and strong
data (measured slopes $1.90$ and $1.89$) and by only $\approx2$ for the
intermediate data (slope $0.96$). The $B$ self--convergence on the finest grid
triple confirms this: $Q=3.75$ (order $1.91$) for the weak data and $Q=2.16$
(order $1.11$) for the intermediate data. Figure~\ref{fig:convres} shows
$\|E\|_2$ versus $\Delta r$ for the three data sets against reference slopes~$1$
and~$2$, and Figure~\ref{fig:convself} shows the canonical scaled--difference
overlay of $B$: the weak difference $B_{h}-B_{h/2}$ coincides with
$4\,(B_{h/2}-B_{h/4})$ (order~$2$), whereas the intermediate one coincides with
$2\,(B_{h/2}-B_{h/4})$ (order~$1$).

\FloatBarrier

\end{document}